\documentclass[authoryear]{elsarticle}  

\usepackage[T1]{fontenc}
\usepackage{lmodern}
\usepackage{amsmath, amssymb, amsthm}
\usepackage{graphicx, booktabs, xcolor}
\usepackage{hyperref}
\hypersetup{colorlinks=false, linkcolor=black, citecolor=blue, urlcolor=blue}
\usepackage{placeins}
\usepackage{multirow}  
\usepackage{booktabs} 
\usepackage{float}

\usepackage[margin=2.54cm]{geometry}

\graphicspath{{./}{./figures/}{./Figures/}{./images/}}
\DeclareGraphicsExtensions{.pdf,.png,.jpg}
\usepackage{subcaption} 

\journal{Nuclear Physics B}

\begin{document}

\begin{frontmatter}


	
	\begin{graphicalabstract}
		\centering
		\includegraphics[width=\linewidth]{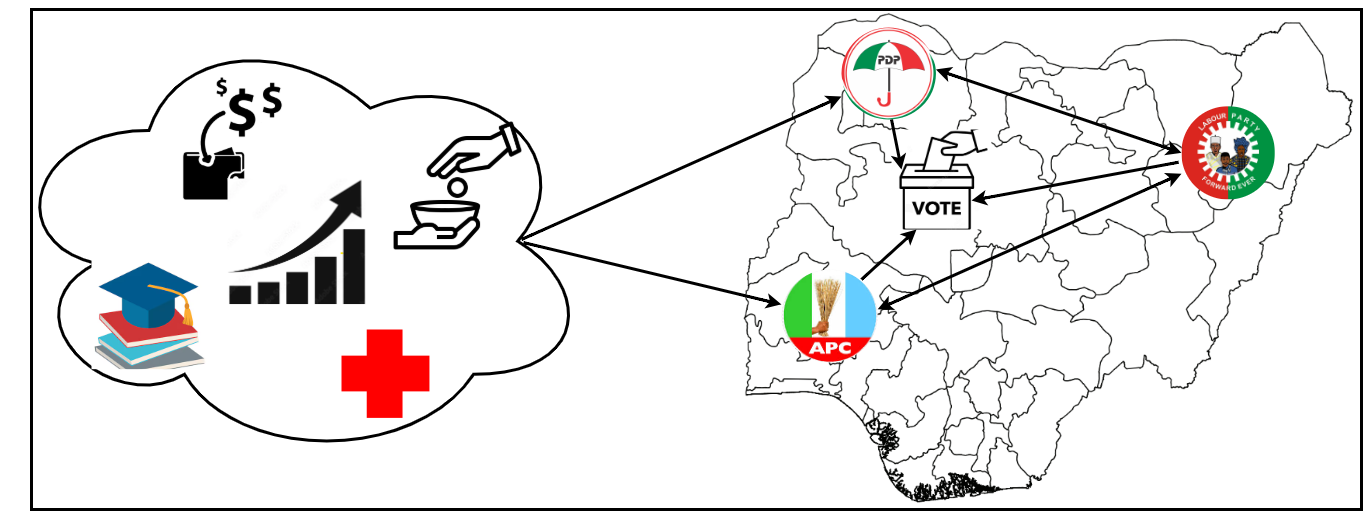} 
	\end{graphicalabstract}

\title{Assessment of Socioeconomic Determinants of Voting Patterns: A Spatio-Temporal and Multinomial Logits Analysis of the 2023 Nigerian Presidential Election}

 \author[1,3]{Vine Nwabuisi Madukpe}
\affiliation[1]{organization={School of Mathematical Sciences, Universiti Sains Malaysia},
             city={Gelugor},
            postcode={11800 USM},
             state={Penang},
             country={Malaysia}}
             
\author[2,3]{Chukwuma Bright Ugoala} 
 \affiliation[2]{organization={Department of Biostatistics, Augusta University},
            addressline={1120 15th Street.},            
            city={Rains Hall},
            postcode={GA 30912},
            state={Augusta},
           country={United States}}

\affiliation[3]{organization={Department of Mathematics, Abia State University},
            city={Uturu},
            postcode={P.M.B. 2000}, 
            state={Abia},
            country={Nigeria}}
            
 \author[4]{Chiemela Uzochi Wokoma}
 \affiliation[4]{
 	organization={Department of Sociology, Abia State University},
 	addressline={P.M.B. 2000},
 	city={Uturu},
 	state={Abia},
 	country={Nigeria}
 }
 
 \author[5]{Gloria Chinenye Nwachukwu}
 \affiliation[5]{
 	organization={Tech4Rurals Foundation},
 	addressline={28 Rumuadaolu Market Road},
 	city={Port Harcourt},
 	postcode={500272},
 	state={River},
 	country={Nigeria}
 }
 

\begin{abstract}
	 In Nigeria, electoral behavior is often interpreted through ethno-religious views and regional allegiances, without empirically assessing the influence of socioeconomic indicators such as health, income, education, and other deprivations on voter behavior. This study investigates the voting pattern in the 2023 Nigerian presidential election and previous cycles using spatio-temporal and multivariate analysis. It examines whether support for some candidates was more substantial in states with higher Human Development Index (HDI) and whether this alignment impacts governance quality and macroeconomic performance post-election. Socioeconomic data were obtained from the Global Data Lab and the Nigerian Bureau of Statistics (NBS) for the year preceding each election, while presidential vote results were sourced from the Independent National Electoral Commission (INEC) portal. The results show that the Labour Party (LP) dominated states with high socioeconomic indices, accounting for about 30–53\% of the variance in the voting patterns of LP and People's Democratic Party (PDP), while the All Progressive Congress (APC) variance is less explained by these factors. Multinomial logits based on the HDI are used to predict party win probabilities; the model predicted about 60\% winning accurately. Additionally, comparative analysis of four presidential election cycles revealed that in 2011, the winner had an HDI-vote correlation of 0.44, with improved macroeconomic indices post-election. In contrast, 2015, 2019, and 2023 saw negative correlations of -0.38, -0.43, and -0.34, respectively, alongside macroeconomic decline post-elections. The findings suggest that socioeconomic development shapes political preferences, promotes issue-based politics, and supports the rise of quality leadership; therefore, strengthening education, healthcare, and poverty reduction should be prioritized to enhance citizens’ well-being and build an informed, reform-oriented electorate.
\end{abstract}


\end{frontmatter}



\section{Introduction}
\label{sec1}
Democracy has long been celebrated as a system that empowers the people, a mechanism through which governance is legitimized by popular consent. Its core advantage lies in its commitment to inclusivity, civil liberties, political accountability, and peaceful transitions of power~\citep{Simsek2025}. The democratic ideal ensures that every citizen, regardless of background, can participate in shaping leadership and public policy. However, democracy is not without its contradictions. In societies marked by significant educational, economic, or informational inequality, democratic processes can be influenced by populism, misinformation, or the targeted persuasion of communities with limited access to information~\citep{Nieminen2024,Duarte2025}. The Nigerian musician and political activist Fela Kuti famously dismissed Africa’s democratic experiments as "demonstrations of craze," criticizing how power often circulates among elites with little accountability to the people~\citep{Oikelome2014}. Also,\citet{Scholtz2024} argued that polarization in Western democracies stems from outdated systems of representation that no longer fit diverse and individualized societies. This mismatch fuels citizen disconnection, party radicalization, and media incentives for division. To counter it, he proposes rethinking representation through a “civil democracy” that restores trust and collective problem-solving. As such, while democracy promises equality, it may, in practice, produce outcomes that reflect the will of the most numerically dominant or easily mobilized, rather than outcomes driven by critical engagement and informed choice.

At the heart of modern democratic practice is the principle of ``one person, one vote'' \citep{Scholtz2024,YaoQin2025} the idea that each citizen’s vote carries equal weight. While this principle reinforces formal equality, it assumes a level playing field in terms of access to information, civic education, and political awareness.Critics have argued that democratic outcomes can suffer when segments of the electorate are disadvantaged by systemic barriers to knowledge or public discourse~\citep{Pottle2024}. This concern is particularly evident in countries like Nigeria, where literacy levels and access to civic information vary widely, and corrupt politicians often exploit these gaps to undermine the tenets of democracy \citep{Alilu2024, Awoyemi2025}. This highlights a broader tension between equality of participation and equity in informed engagement, raising the difficult and a controversial question "Should all votes carry the same weight when citizens differ so markedly in their access to reliable political information?"
 While this question might appear controversial, it raises legitimate concerns about the role of informational equity in sustaining healthy democratic processes. Scholars such as Jason Brennan have explored epistocracy — rule by the knowledgeable — as a theoretical response to these concerns, though the idea remains contentious~\citep{Lucky2023, Keeling2025}. However, \citet{Manor2025} proposes a hybrid model that combines local epistocracy with national democracy as a compromise between expertise and democratic legitimacy. Local governance is seen as better suited for epistocracy because of tighter feedback, room for experimentation, and potential policy gains, while posing fewer risks to democratic identity.

The tension between the quantity and quality of votes is not merely theoretical, but finds practical expression in existing institutions. The United States’ electoral system, for example, incorporates the Electoral College — a mechanism designed to balance popular votes with regional representation, thereby tempering pure majoritarian rule. While the system has been criticized for distorting the popular will, as seen in the 2000 and 2016 elections, it is also viewed by some as a safeguard against purely numerical dominance~\citep{AshCenter2025}. The Electoral College illustrates how democracies might incorporate balancing mechanisms not based on education or wealth, but on geography to protect minority interests. Although Nigeria employs a form of geographical balancing similar to the United States, requiring a candidate to secure at least 25\% of the vote in two thirds of the 36 states and the Federal Capital Territory to be declared the winner, however, the US model differ. Both systems highlight enduring concerns about how best to balance democratic practice with structural adjustments for disparities in education, media access, socioeconomic conditions, or civic awareness \citep{Johnson2024}. These concerns raise critical questions about whether democratic processes reward mere numerical advantage over informed deliberation, and what this means for the legitimacy and effectiveness of elected leadership.

Public philosophies and academic research on democracy have long recognized these dilemmas. John Stuart Mill warned in the 19th century that universal suffrage without universal education risks producing decisions driven more by habit than by critical reasoning~\citep{Johnson2024}. Modern studies confirm that political misinformation spreads more readily in contexts where access to quality education and media literacy is limited~\citep{AdjinTettey2022,GuessMunger2022}. While democracy aspires to reflect the will of the people, it may, in practice, reflect the influence of those who can most easily shape public opinion. The universal suffrage is foundational to democratic legitimacy, however, equal voting power in unequal societies may produce outcomes that do not reflect the most accountable or capable leadership. A candidate with access to expansive patronage networks or media platforms may be able to consolidate support in areas with limited education, socioeconomic well-being, civic infrastructure and defeat another candidate whose appeal lies primarily in more informed or critically engaged constituencies. This paradox has been insufficiently studied in African democracies. Most electoral research focuses on qualitative analysis of turnout, ethno-religious alignment, or party loyalty, rather than quantitative on the relationship between voter access to development, resources and electoral outcomes~\citep{Nwangbo2024,Fubara2025}. 

Beyond theoretical debates on the impact of quality or quantity of vote in democracy across the world and in particular Nigeria, scholars analyzing 2023 Nigerian presidential election emphasize that the country’s democratic processes are shaped by weak party ideology, ethno-religious cleavages, and social media mobilization and misinformation reiterating the concern of inequalities. \citet{Babalola2024} argued that Nigerian political parties no longer function as ideological platforms but as vehicles for patronage and elite bargaining. Consequently, electoral competition is less about policy vision and but more about access to resources and coalition building, producing what he terms a "democratic deficit". \citet{IdowuIyabode2024} similarly opined that voting in 2023 largely followed entrenched ethno-religious and regional lines, with northern Muslim-majority states favoring the All Progressives Congress (APC), while southern and Christian-dominated areas leaned toward the Labour Party (LP) and the People’s Democratic Party (PDP). These findings suggest that Nigerian democracy remains constrained by identity politics, limiting the growth of issue-based engagement.  

Other studies highlight new democratic dynamics, particularly the role of social media and youth mobilization.  \citet{Olabanjo2023} and \citet{Oyewola2023} demonstrate that Twitter and other platforms amplified Peter Obi’s reformist image and broadened youth participation. Analyses of millions of tweets revealed that Obi attracted the most positive impressions online, while Tinubu and Atiku relied more on traditional patronage networks. \citet{Salahu2023} adds that religion remained decisive, with Christian and Muslim blocs consolidating around different candidates, reinforcing Nigeria’s persistent electoral fault lines. In all, these studies show that the 2023 election reflects both continuity and change: continuity in identity-driven voting, and change in the growing influence of digital mobilization and demands for accountability by younger, urban voters. Despite this extensive focus on ethno-religious factors, to the best of the author’s knowledge no empirical study has examined how multidimensional poverty, health quality, income, and literacy level shaped voting behavior across Nigerian states in the 2023 presidential election. This gap underscores the need for a broader analysis that integrates socioeconomic development indicators alongside and the voting pattern observed in 2023 which this study aims to address.

Therefore, the purpose of this study is to empirically examine how multidimensional poverty, health quality, income, and educational status influenced voting patterns in the 2023 Nigerian presidential election. The contest was highly competitive among three major candidates with distinct political and professional backgrounds: Peter Obi of the LP, often regarded as a reform-oriented technocrat; Atiku Abubakar of the PDP, a veteran politician with extensive federal experience; and Bola Ahmed Tinubu of the APC, a dominant figure within Nigeria’s political establishment. Understanding how socioeconomic realities shaped support for these contenders requires a structured analysis, which this study provides. To achieve this aim, the study is organized as follows: The first section outlines the data sources and methodology used in the analysis. The second section presents the spatio-temporal distribution of voting patterns across Nigerian states in relation to socioeconomic variables such as health, income, education, and other deprivations captured multidimensional poverty. The third section models party win probabilities based on these socioeconomic indicators. The fourth section employs univariate analyses to compare the correlation between socioeconomic factors, voting patterns, and macroeconomic outcomes across electoral cycles. The final section discusses the broader implications of these findings for democratic theory, electoral equity, and policy reform. Ultimately, this study contributes empirically to debates about whether, beyond ethno-religious factors, socioeconomic conditions also shape voting behavior in Nigeria and whether the socioeconomic profile of voters influences the quality of elected leadership. This research therefore seeks to address the following research questions (RQ):

RQ1: How do spatial and variations in income, health quality, education, and other deprivation captured by multidimensional poverty across 36 states of Nigeria including the the federal capital territory describe the voting patterns observed in the 2023 presidential election?

RQ2: To what extent do socioeconomic development indicators specifically income, health quality, educational status, and their composite with other deprivation in the form of  multidimensional poverty quantitatively explain voting support for the three major political parties (LP, APC, and PDP) in the 2023 presidential election?

RQ3: Does the HDI, as an aggregate of socioeconomic indicators, influenced the likelihood that a state is won by the LP, PDP, or APC in the 2023 presidential election?

RQ4: Is a stronger positive correlation between state-level HDI and votes for the eventual winner associated with better macroeconomic outcomes during that tenure, compared to electoral cycles with weaker or negative correlations?

\section{Data and Methods}

\subsection{Study Area}
\begin{figure}[htbp]
	\centering
	\includegraphics[width=0.9\linewidth]{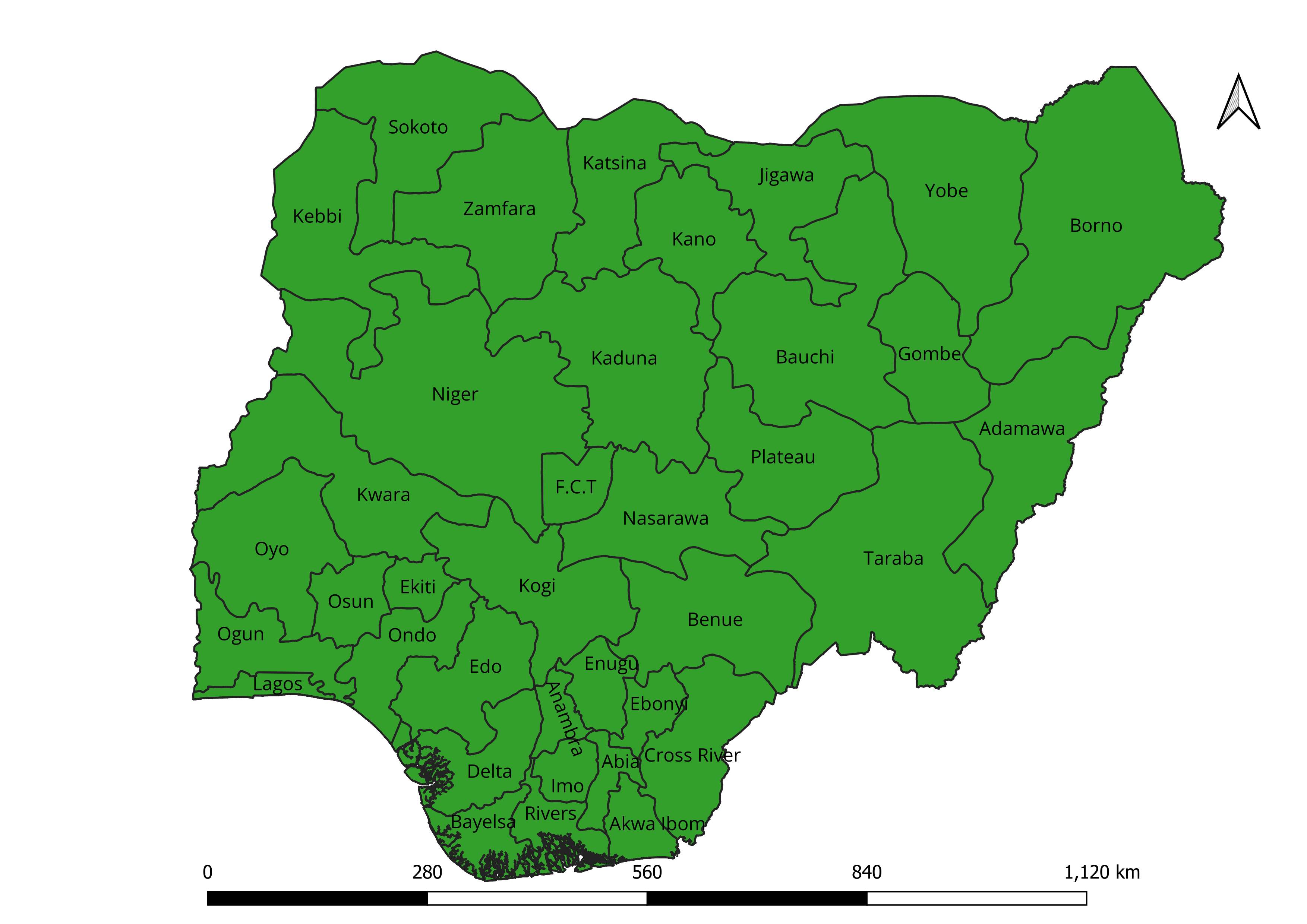} 
	\caption{Study area: Nigeria showing the 36 states and the Federal Capital Territory (F.C.T)}
	\label{fig:study-area}
\end{figure}

Nigeria, a West African federation and the most populous Black nation, have an estimated population of 223.8 million in 2023; it covers about 923,768 km$^{2}$ and is centered near $10^{\circ}$N, $8^{\circ}$E \citep{WorldBank2023}. Administratively, it comprises 36 states and the Federal Capital Territory (F.C.T) as shown in Figure~\ref{fig:study-area}. The 2023 presidential election was conducted on 25~February~2023 by the Independent National Electoral Commission (INEC), with 93{,}469{,}008 registered voters and 176{,}846 planned polling units nationwide. These geographic and electoral characteristics make Nigeria an appropriate setting for state-level spatio-temporal and multivariate analysis of voting behavior; the next subsection presents the datasets and variables employed in the study.

	\FloatBarrier

\subsection{Dataset}

The data analyzed in this study are secondary sources drawn from the Global Data Lab (GDL)\citep{GlobalDataLab2022} to ensure comparability of socialeconomics variables across the 36 states of Nigeria and F.C.T. GDL is a reputable database aligned with international definitions used by the UNDP; it publishes detailed documentation for the Subnational Human Development Index (SHDI) and maintains national averages consistent with the UNDP series \citep{SmitsPermanyer2019}. The variables explored in this study are: the Income Index (II), a measure of material living standards derived from gross national income per capita in purchasing power parity terms; the Health Index (HI), a summary of health attainment proxied by life expectancy at birth; the Education Index (EI), an education attainment measure combining mean years of schooling for adults and expected years of schooling for children and their composite the HDI were extracted for 2022, 2019, 2014, and 2010 for all 36 states and the F.C.T, providing the most recent pre election year data for each presidential election cycle. These indicators jointly offer a consistent empirical profile of human development and socioeconomic conditions. Additionally, Nigerian National Multidimensional Poverty Index (MPI), which measured a wider spectrum of deprivation such as access to sanitation, clean energy, housing quality, and exposure to shocks that are not explicitly reflected in the stand alone indices is equally obtained. It comprises 15 indicators across four dimensions: Health (Nutrition, Child Mortality), Education (Years of Schooling, School Attendance), Living Standards (Cooking Fuel, Sanitation, Drinking Water, Electricity, Housing, Assets), and Work \& Shocks (Employment, Security, Time-use, Shock Exposure) \citep{nbsnmpi2022,dattoma2023multidimensional}. Finally, complementary electoral data for the 2023, 2019, 2015, and 2011 presidential elections, covering vote totals by state for the LP, PDP, APC, and other parties, were obtained from the official INEC results website, ensuring a verifiable record of voting patterns.

\subsection{Methodologies (Spatio–Temporal, Correlation, and Multinomial)}

For the spatio–temporal analysis of voting patterns in the 2023 Nigerian presidential election across the 36 states and the F.C.T, we employ QGIS~v3.44 \citep{QGIS_software} to integrate and visualize spatial layers, mapping the II, HI, MPI, and EI alongside electoral outcomes for LP, APC, and PDP. According to the United Nations Development Programme (UNDP), HDI is calculated as \citep{desiqueira2022undp}:
\begin{equation}
	HDI = \left( {HI} \cdot {EI} \cdot {II} \right)^{\tfrac{1}{3}}
	\label{eq:hdi}
\end{equation}
 
MPI is inverted such that all indices (II, HI, EI, MPI) move in the same direction, higher values consistently mean higher development and better conditions given as:
\begin{equation}\label{eq:invmpi}
	\text{Inverted\_MPI}_s \;=\; \frac{100 - \text{MPI}_s}{100}\,,
\end{equation}

\subsubsection{Spearman’s rank correlation.}
To quantify the association between socioeconomic variables and party performance, we correlate each state’s HDI with the votes received by each party using Spearman’s rank correlation \citep{Michelucci2025}:
\begin{subequations}\label{eq:spearman}
	\begin{equation}\label{eq:spearman_ranks}
		\rho=\frac{\sum_{i=1}^{n}(R_{X,i}-\bar R_X)(R_{Y,i}-\bar R_Y)}
		{\sqrt{\sum_{i=1}^{n}(R_{X,i}-\bar R_X)^2}\,\sqrt{\sum_{i=1}^{n}(R_{Y,i}-\bar R_Y)^2}},
	\end{equation}
	where $R_{X,i}$ and $R_{Y,i}$ are the ranks of HDI and votes in state $i$ (average ranks are used for ties). In the special case of no ties, the equivalent form is
	\begin{equation}\label{eq:spearman_noties}
		\rho=1-\frac{6\sum_{i=1}^{n} d_i^{2}}{n\,(n^{2}-1)}, \qquad d_i=R_{X,i}-R_{Y,i}.
	\end{equation}
\end{subequations}

\subsubsection{Multinomial logistic regression (MNL)}
We model the probability that a party wins state \(s=1,\ldots,37\) using a multinomial logit with the HDI \citep{Swami2025}.
Let \(y_s\in\{\text{APC},\text{LP},\text{PDP}\}\) and define party–specific linear predictors
\begin{equation}
	\eta_{p,s} \;=\; \alpha_p \;+\; \beta_p\, \text{HDI}_s
	\qquad p\in\mathcal{P}=\{\text{APC},\text{LP},\text{PDP}\}.
	\label{eq:mnl-linear}
\end{equation}
Winning probabilities of each party follow the softmax form:
\begin{equation}
	\Pr(y_s=p \mid \text{HDI}_s)
	\;=\; \frac{\exp(\eta_{p,s})}{\sum_{q\in\mathcal{P}}\exp(\eta_{q,s})}.
	\label{eq:mnl-prob}
\end{equation}
For identification, APC is taken as the reference and set \(\eta_{\text{APC},s}\equiv 0\), note either of the parties can be chosen as a reference. The model Equation~\eqref{eq:mnl-linear} can then be expressed as log–odds relative to APC:
\begin{align}
	\log\frac{\Pr(y_s=\mathrm{LP})}{\Pr(y_s=\mathrm{APC})}
	&= \alpha_{\mathrm{LP}} + \beta_{\mathrm{LP}}\,\mathrm{HDI}_s, \label{eq:mnl-odds-lp}\\
	\log\frac{\Pr(y_s=\mathrm{PDP})}{\Pr(y_s=\mathrm{APC})}
	&= \alpha_{\mathrm{PDP}} + \beta_{\mathrm{PDP}}\,\mathrm{HDI}_s. \label{eq:mnl-odds-pdp}
\end{align}

\begin{itemize}
	\item $s$: state index (36 states $+$ F.C.T).
	\item $p$: party index.
	\item $\eta_{p,s}$: linear predictor (systematic utility / log-odds) for party $p$ in state $s$.
	\item $\alpha_p$: party-specific intercept (baseline support for $p$).
	\item $\beta_p$: effect of state $\mathrm{HDI}$ on support for party $p$ (party-specific slope).
\end{itemize}

The coefficient \(\beta_p\) measures how a one–unit increase in HDI given in Equation~\eqref{eq:csds} changes the log–odds that party \(p\) wins relative to APC; \(\exp(\beta_p)\) is the corresponding odds ratio.
Parameters are estimated by maximum likelihood, and inference uses heteroskedasticity–robust (HC3) standard errors.
The MNL assumes independence of irrelevant alternatives (IIA); robustness checks assess this assumption and alternative specifications.

\section{Results}
\subsection{2023 Presidential Election Spatial Analysis}
\begin{figure}[htbp]
	\centering
	
	\begin{subfigure}[t]{0.8\linewidth}
		\centering
		\includegraphics[width=\linewidth]{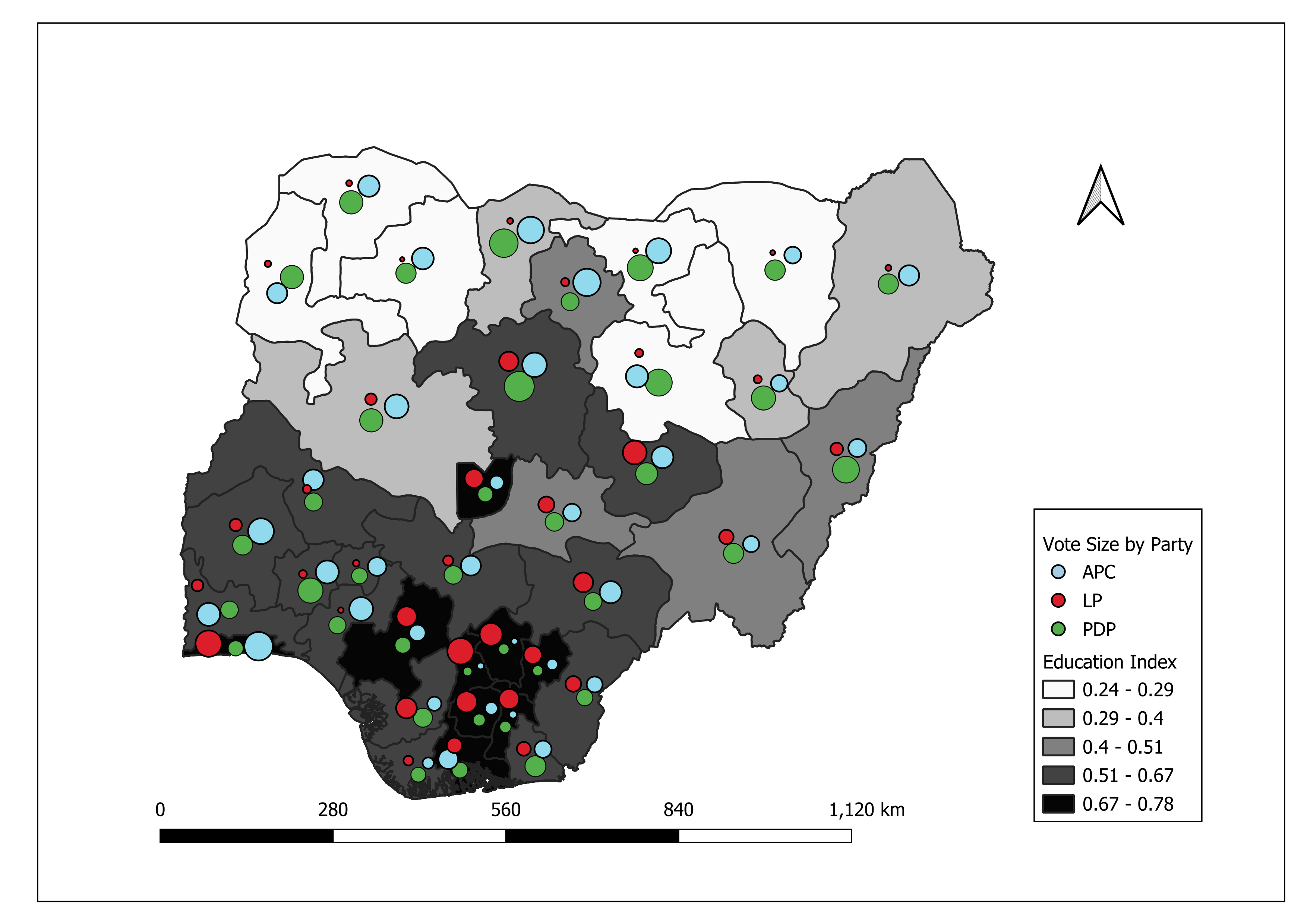} 
		\caption{}
		\label{fig:(a)}
	\end{subfigure}\hfill
	%
	\begin{subfigure}[t]{0.8\linewidth}
		\centering
		\includegraphics[width=\linewidth]{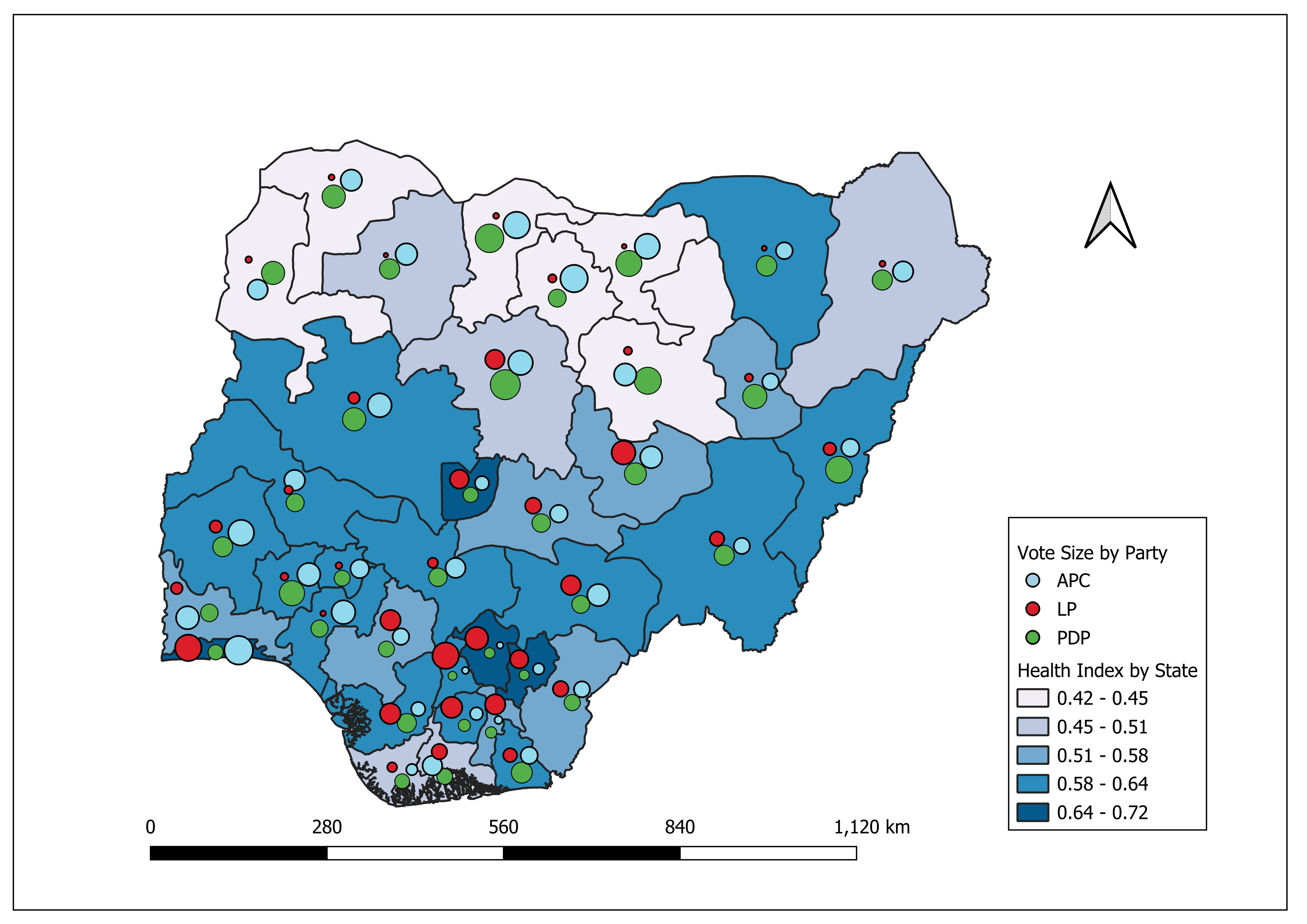} 
		\caption{}
		\label{fig:(b)}
	\end{subfigure}
	
	\caption{State-level 2022 (a) Education Index (b) Health Index and party total votes in the 2023 Nigerian presidential election: }
	\label{fig:fig2}
\end{figure}

To address RQ1, how variations in II, HI, EI, and MPI across the 36 states and the F.C.T might have shaped voting patterns in the 2023 presidential election, a spatial analysis of 2022 EI,II, HI, and MPI in relation to LP, APC, and PDP vote totals is performed for all states and the F.C.T. Figures~\ref{fig:fig2} an 3 present proportional-symbol choropleths: states are shaded based on the socioeconomic variables, and overlaid circles are colored based on the parties with their size proportional to state vote totals. This visualization supports a qualitative assessment of state voting variation , as well as their spatial correspondence with party support, to quantify this relationships, the subsequent sections provides scatter plots showing the relationship between the socioeconomic variables and the votes recorded by each party. 

\begin{figure}[htbp]
	\centering
	
	\begin{subfigure}[t]{0.8\linewidth}
		\centering
		\includegraphics[width=\linewidth]{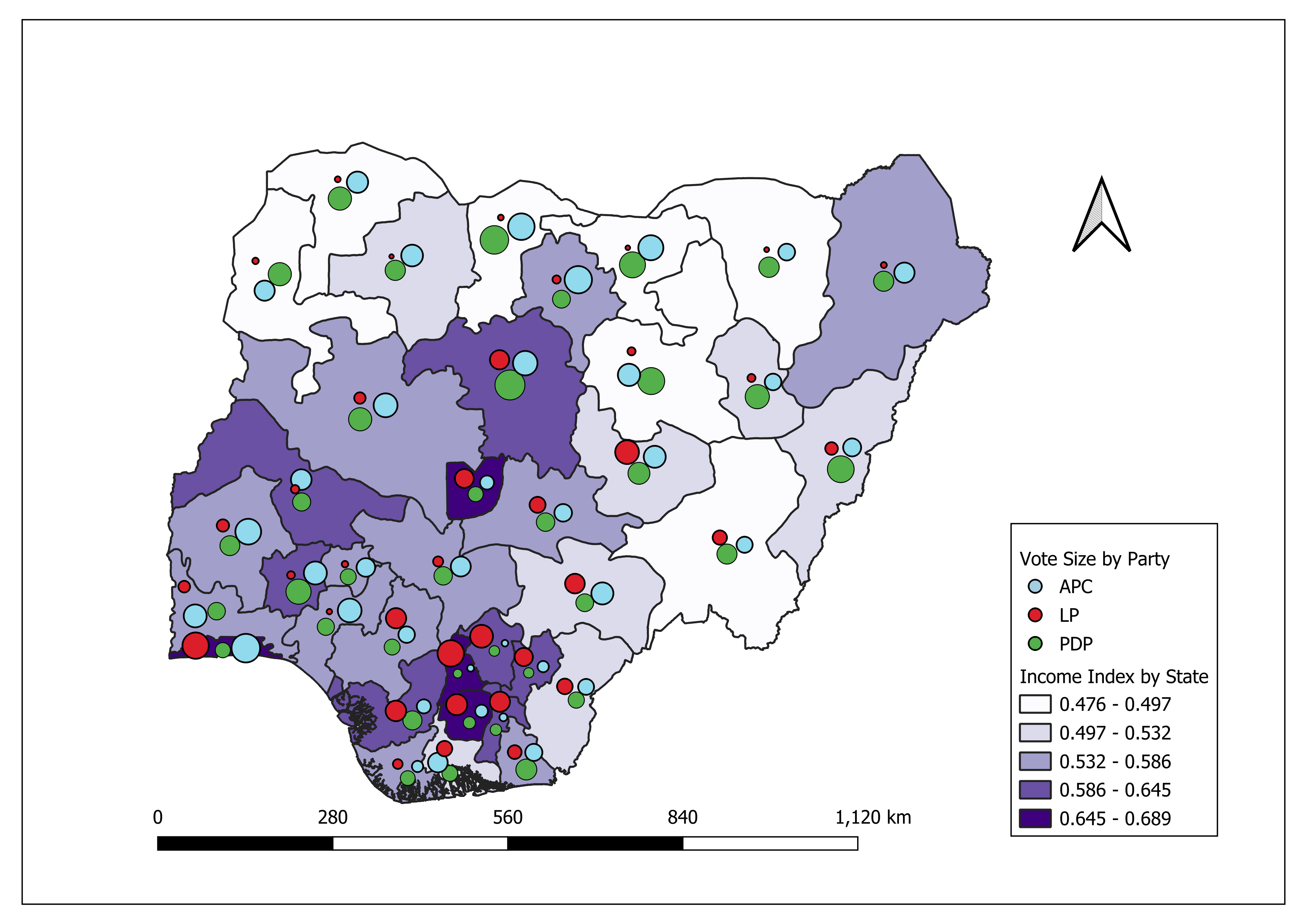} 
		\caption{}
		\label{fig:(3a)}
	\end{subfigure}\hfill
	%
	\begin{subfigure}[t]{0.8\linewidth}
		\centering
		\includegraphics[width=\linewidth]{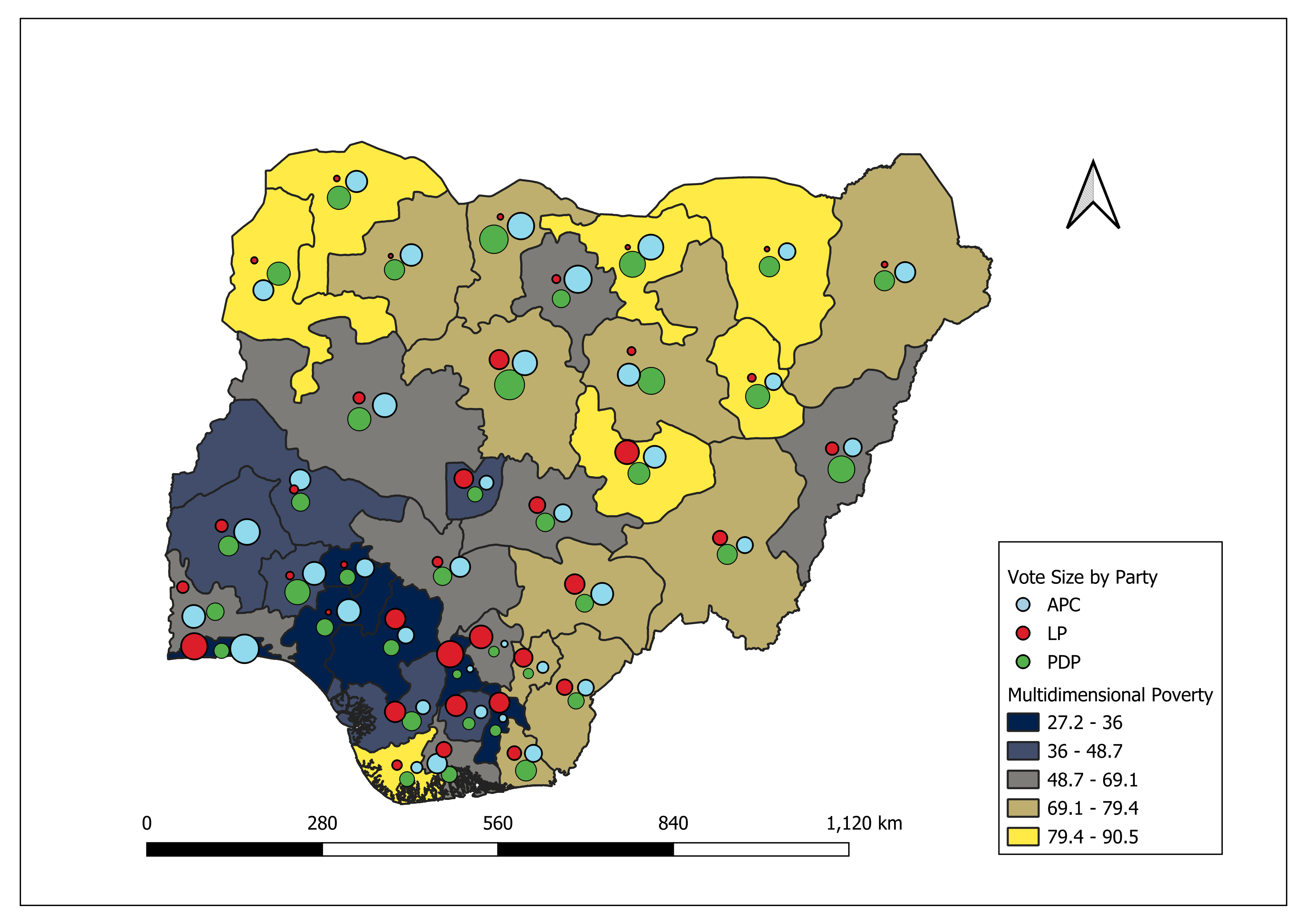} 
		\caption{}
		\label{fig:(3b)}
	\end{subfigure}
	
	\caption{State-level 2022 (a) Income Index (b) Multidimensional Poverty Index and party total votes in the 2023 Nigerian presidential election}
	\label{fig:fig3}
\end{figure}

Figure~\ref{fig:(a)} presents a state-level assessment of the Education Index alongside party total votes. Among the ten states with the highest EI, LP leads in nine—spanning the South-East, South-South, South-West, and North-Central subregions—while APC leads in a single South-South state; PDP does not lead within this subset. This pattern indicates a strong positive association between higher educational attainment and LP support, with comparatively limited traction for APC and PDP in high-EI contexts. To assess whether a similar alignment holds for broader economic well-being, Figure~3 examines the Income Index. In Figure~\ref{fig:(a)} focuses on the HI and shows a broadly similar voting pattern, where LP won six out of the top ten states with highest HI, whereas APC and PDP secured three and one top ten states respectively. The results suggest that states with stronger education and health indices voted LP more compared with other parties. To examine whether this alignment extends to broader economic well-being, Figure~\ref{fig:fig3} analyzes the income and multidimensional poverty of the electorates with respect their voting pattern.

Income and multidimensional poverty are spatially analyzed and presented in Figure~\ref{fig:fig3}. Out of the top ten states with the highest II, LP wins six, APC three, and PDP one. In Figure~\ref{fig:(3a)} and Figure~\ref{fig:(3b)}, among the top ten states with the lowest MPI, LP wins seven, APC two, and PDP one. Electorates in states with higher EI, II, and HI, together with lower MPI, tended to lean toward LP. These visual patterns are further quantitatively analyzed in the subsequent section using correlation and multivariate regression.

\vspace{-1.0em}
\FloatBarrier
\vspace{-2.0em}

\FloatBarrier

\subsubsection{Relationship between Party Votes and the Socioeconomic Indicators}

\begin{figure}[htbp]
	\centering
	
	\begin{subfigure}{0.5\textwidth}
		\includegraphics[width=\linewidth]{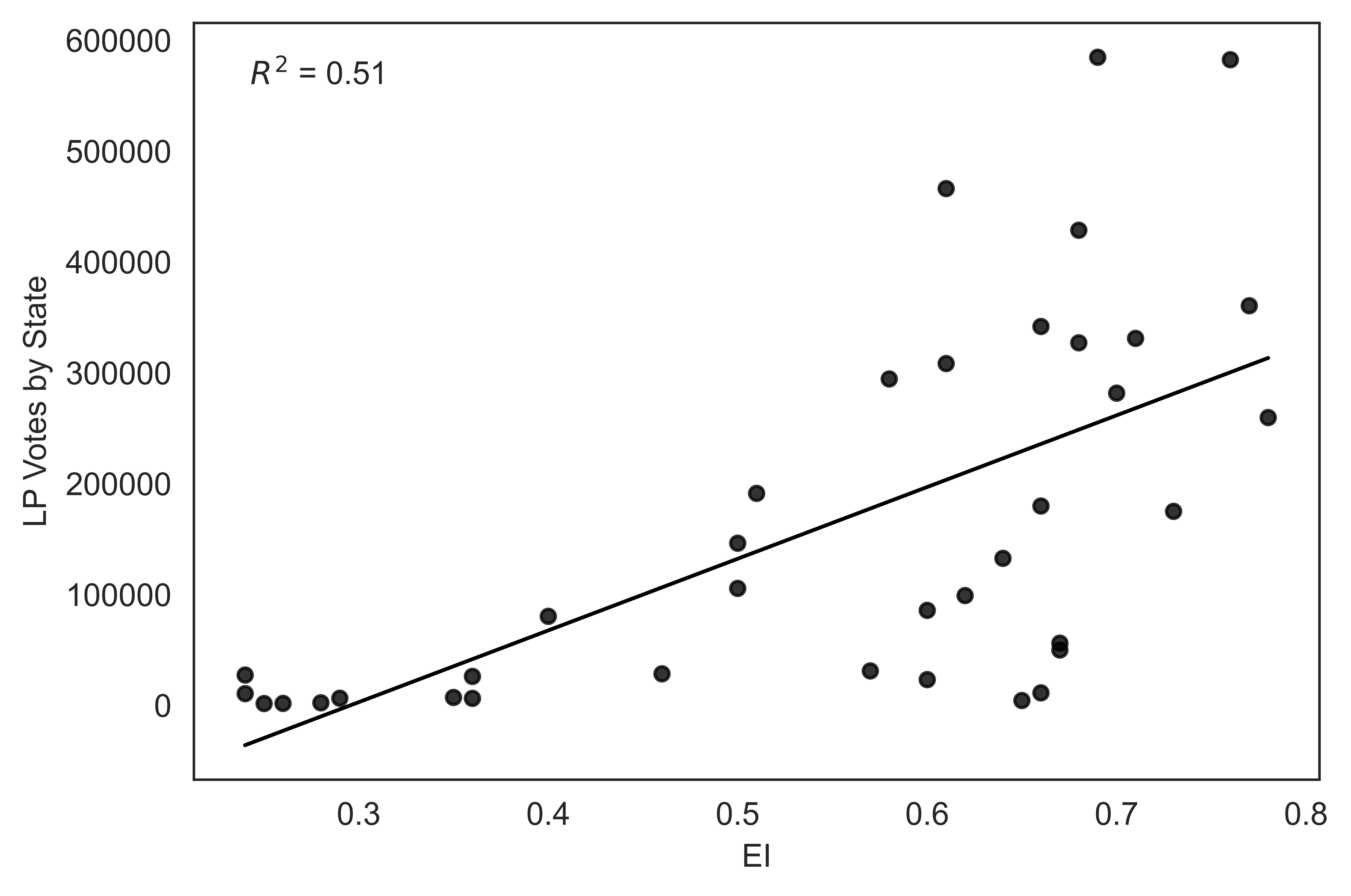}
		\caption{}\label{fig:lp-ei}
	\end{subfigure}\hfill
	\begin{subfigure}{0.5\textwidth}
		\includegraphics[width=\linewidth]{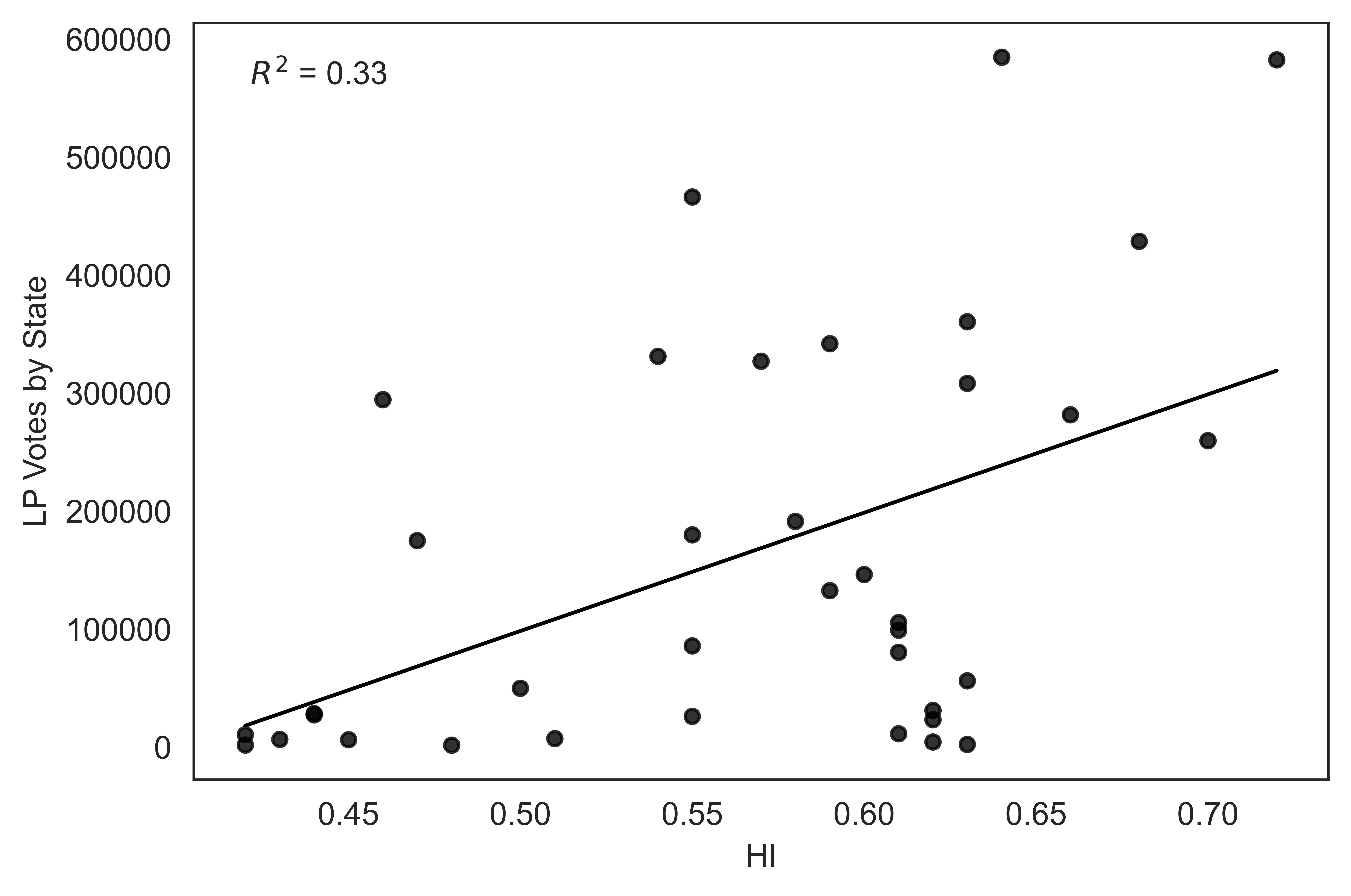}
		\caption{}\label{fig:lp-hi}
	\end{subfigure}
	
	\vspace{0.5em}
	
	\begin{subfigure}{0.5\textwidth}
		\includegraphics[width=\linewidth]{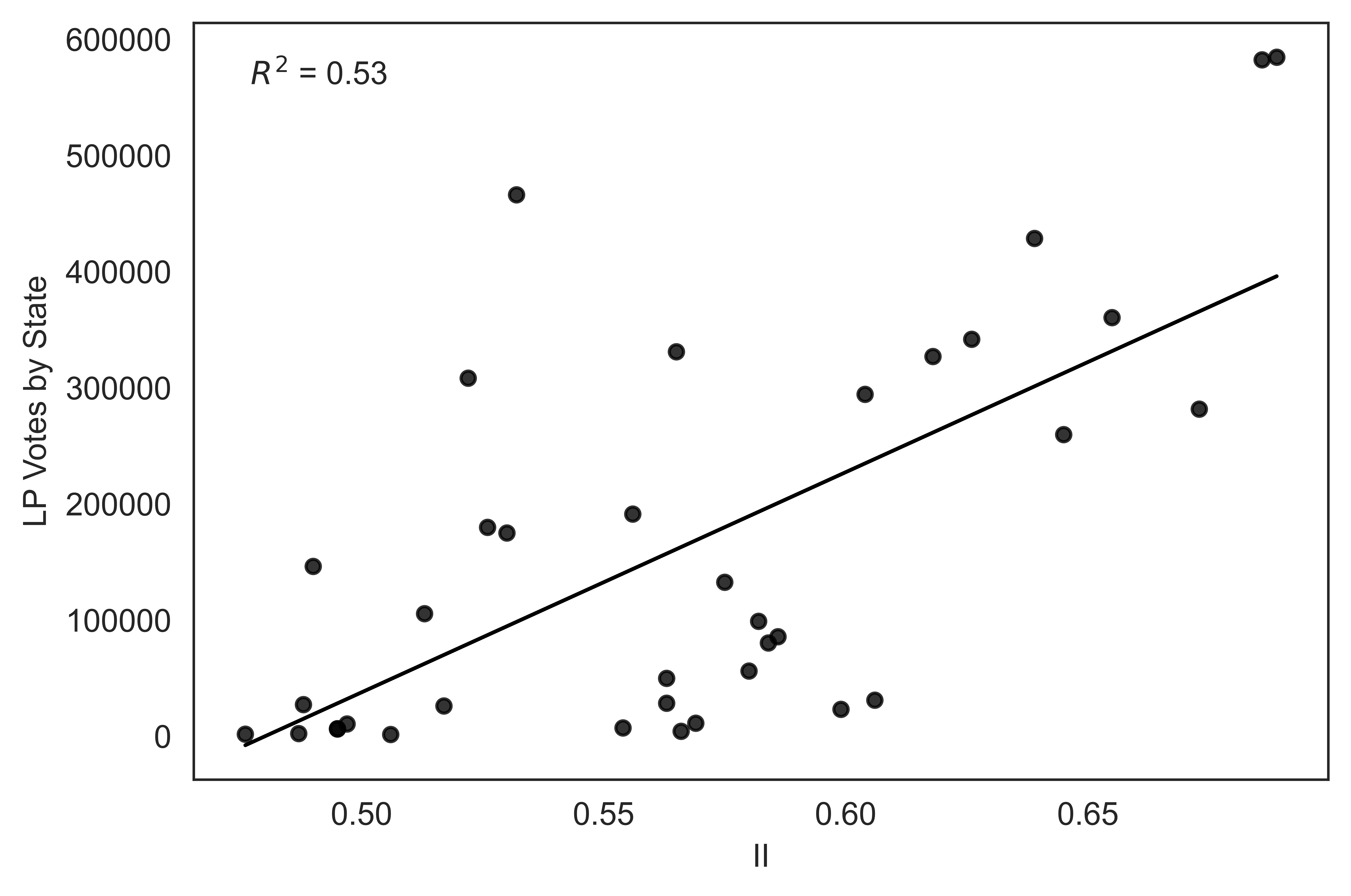}
		\caption{}\label{fig:lp-ii}
	\end{subfigure}\hfill
	\begin{subfigure}{0.5\textwidth}
		\includegraphics[width=\linewidth]{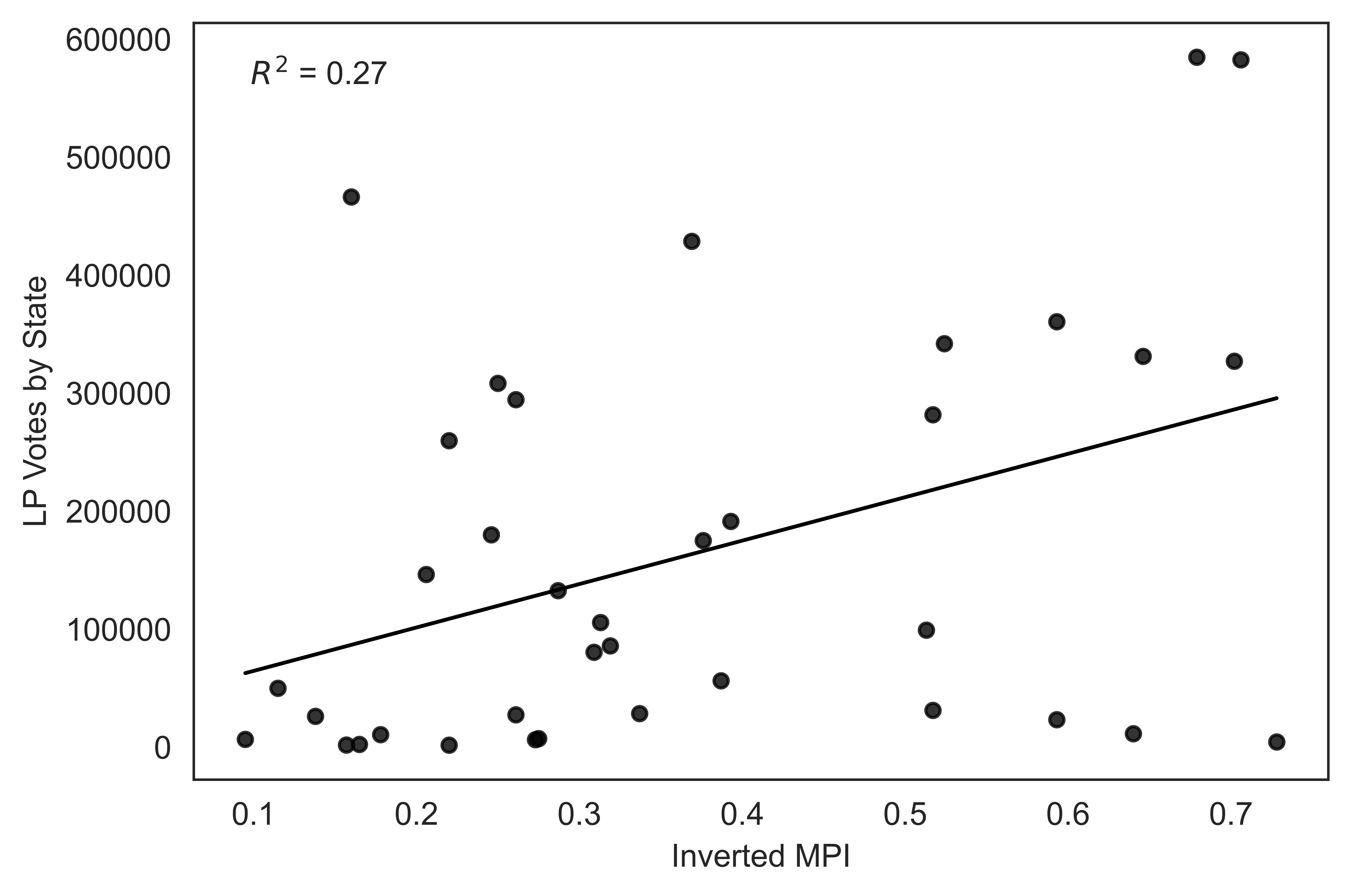}
		\caption{}\label{fig:lp-invmpi}
	\end{subfigure}
	
	\caption{LP votes versus socioeconomic indicators.}
	\label{fig:lp-indicator}
  \end{figure}

Figure~\ref{fig:lp-indicator} illustrates the relationship between LP votes and selected
socioeconomic indicators. The results reveal consistently positive associations
between LP support and measures of socioeconomic development, though with varying
explanatory power.In Figures~\ref{fig:lp-ei}, the EI shows a moderately strong 
positive relationship, explaining about 51\% of the variation in LP votes, 
indicating that states with higher educational attainment tended to give more 
votes to the LP. Figure~\ref{fig:lp-hi} highlights an equally strong positive association 
between the II and LP votes, accounting for about 53\% of the variation, 
suggesting that wealthier states disproportionately supported the LP. By contrast, 
Figures~\ref{fig:lp-ii} and \ref{fig:lp-invmpi} suggest that states with improved health quality and lower poverty levels were more likely to favour the LP, although these indicators 
explained a smaller share of the variation. Overall, these results suggest that support for the LP in the 2023 election was stronger in areas with higher levels of socioeconomic development. However, other factors may have contributed to the unexplained variation in voting behaviour. To complement this analysis, Figure~5 presents the corresponding relationships for the APC, enabling a comparative understanding of how socioeconomic factors influenced voting patterns across major parties.

	\FloatBarrier

\begin{figure}[htbp]
	\centering
	
	\begin{subfigure}{0.5\textwidth}
		\includegraphics[width=\linewidth]{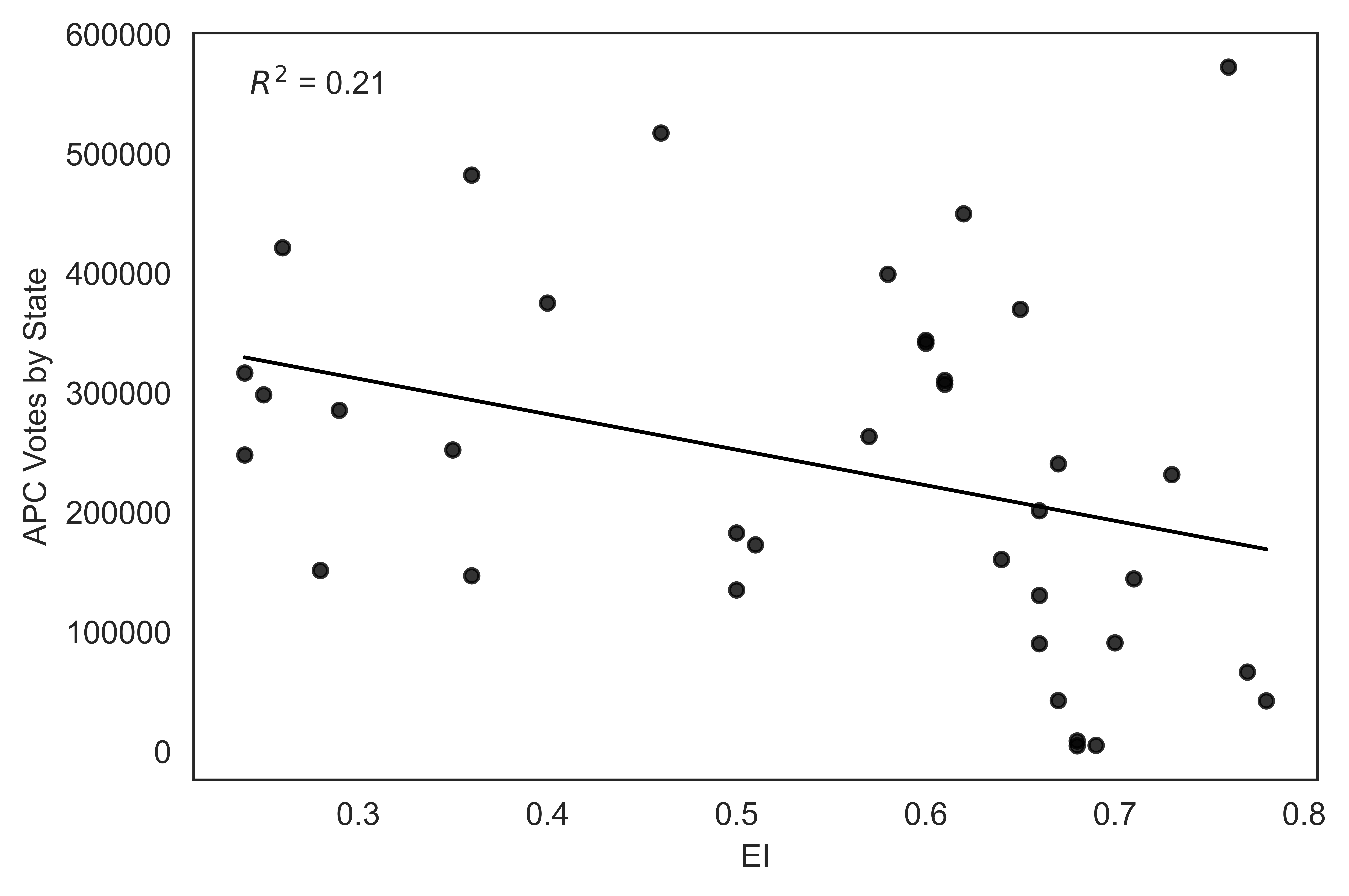}
		\caption{}\label{fig:apc-ei}
	\end{subfigure}\hfill
	\begin{subfigure}{0.5\textwidth}
		\includegraphics[width=\linewidth]{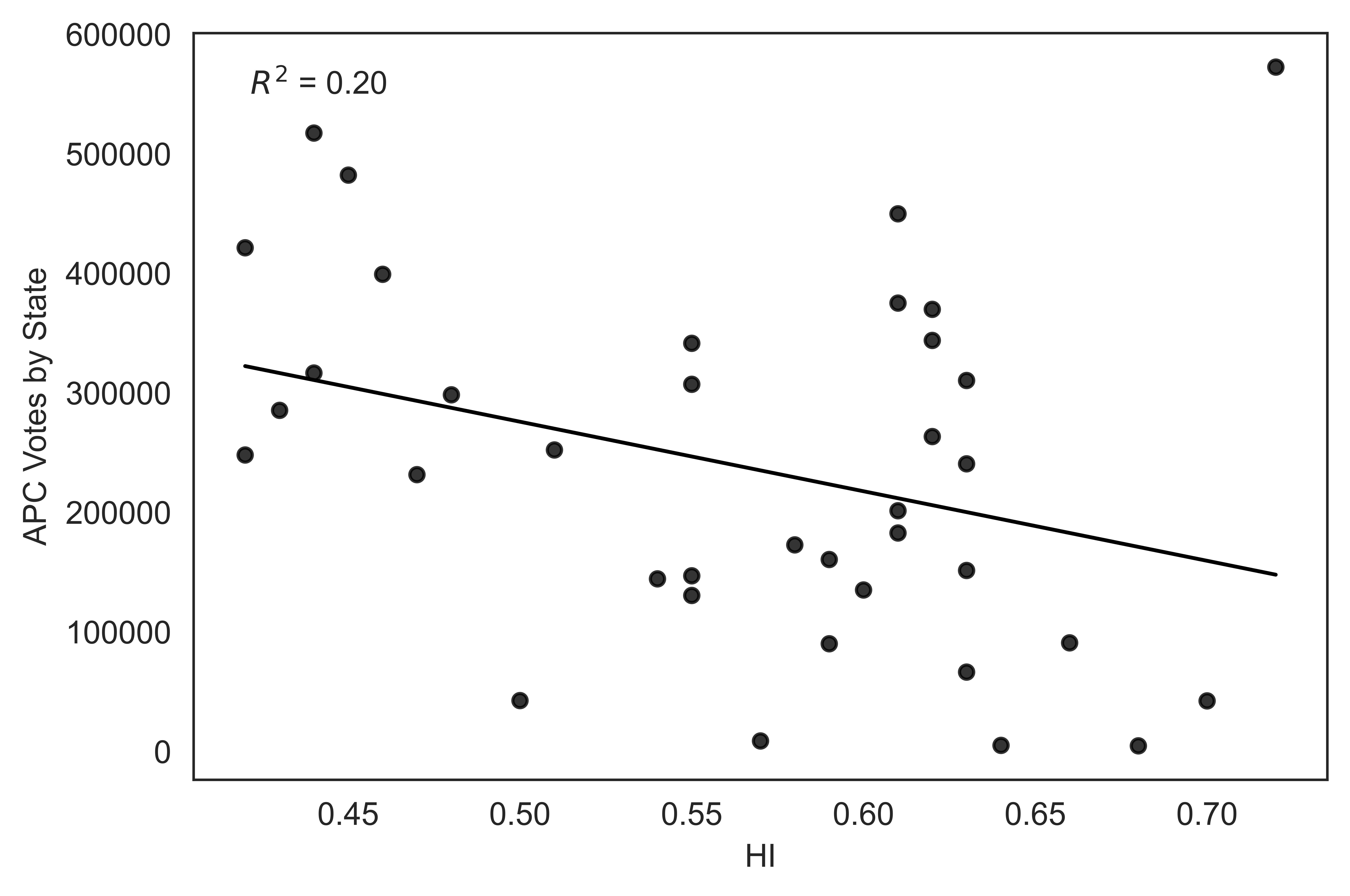}
		\caption{}\label{fig:apc-hi}
	\end{subfigure}
	
	\vspace{0.5em}
	
	\begin{subfigure}{0.5\textwidth}
		\includegraphics[width=\linewidth]{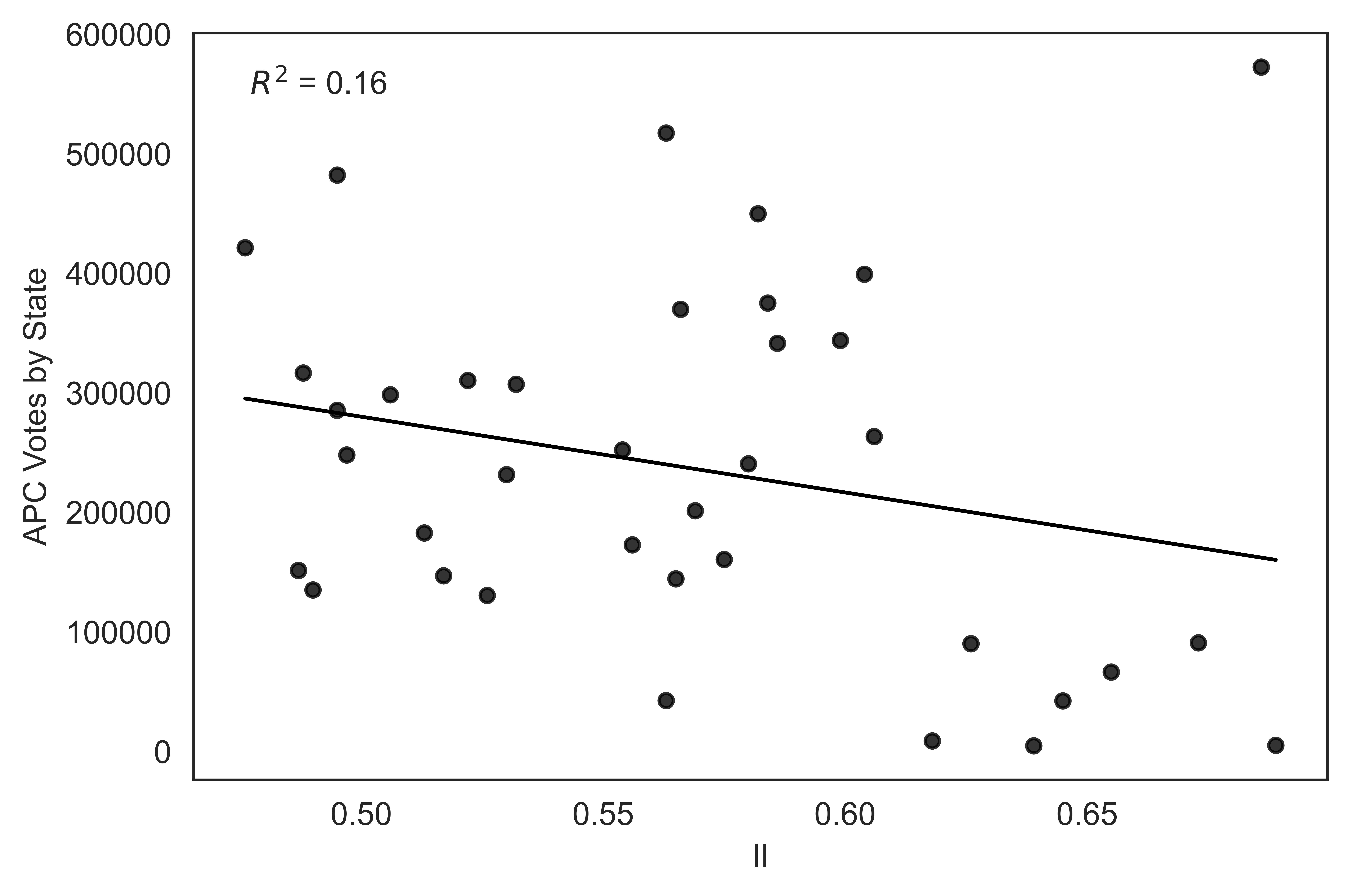}
		\caption{}\label{fig:apc-ii}
	\end{subfigure}\hfill
	\begin{subfigure}{0.5\textwidth}
		\includegraphics[width=\linewidth]{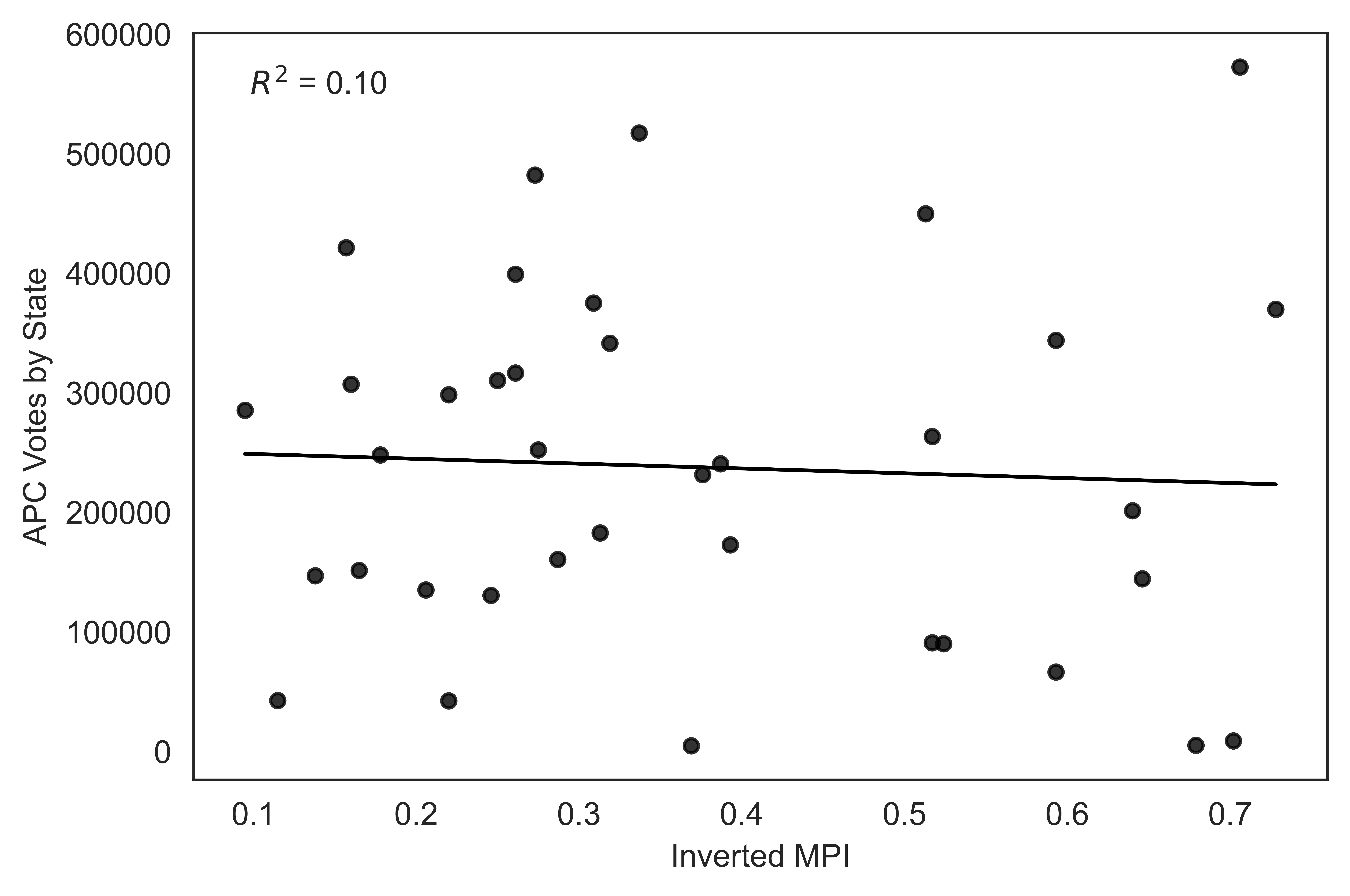}
		\caption{}\label{fig:apc-invmpi}
	\end{subfigure}
	
	\caption{APC votes versus socioeconomic indicators.}
	\label{fig:apc-indicators}
\end{figure}

\begin{figure}[H] 
	\vspace*{-2em}
	\centering
	
	\begin{subfigure}{0.5\textwidth}
		\includegraphics[width=\linewidth]{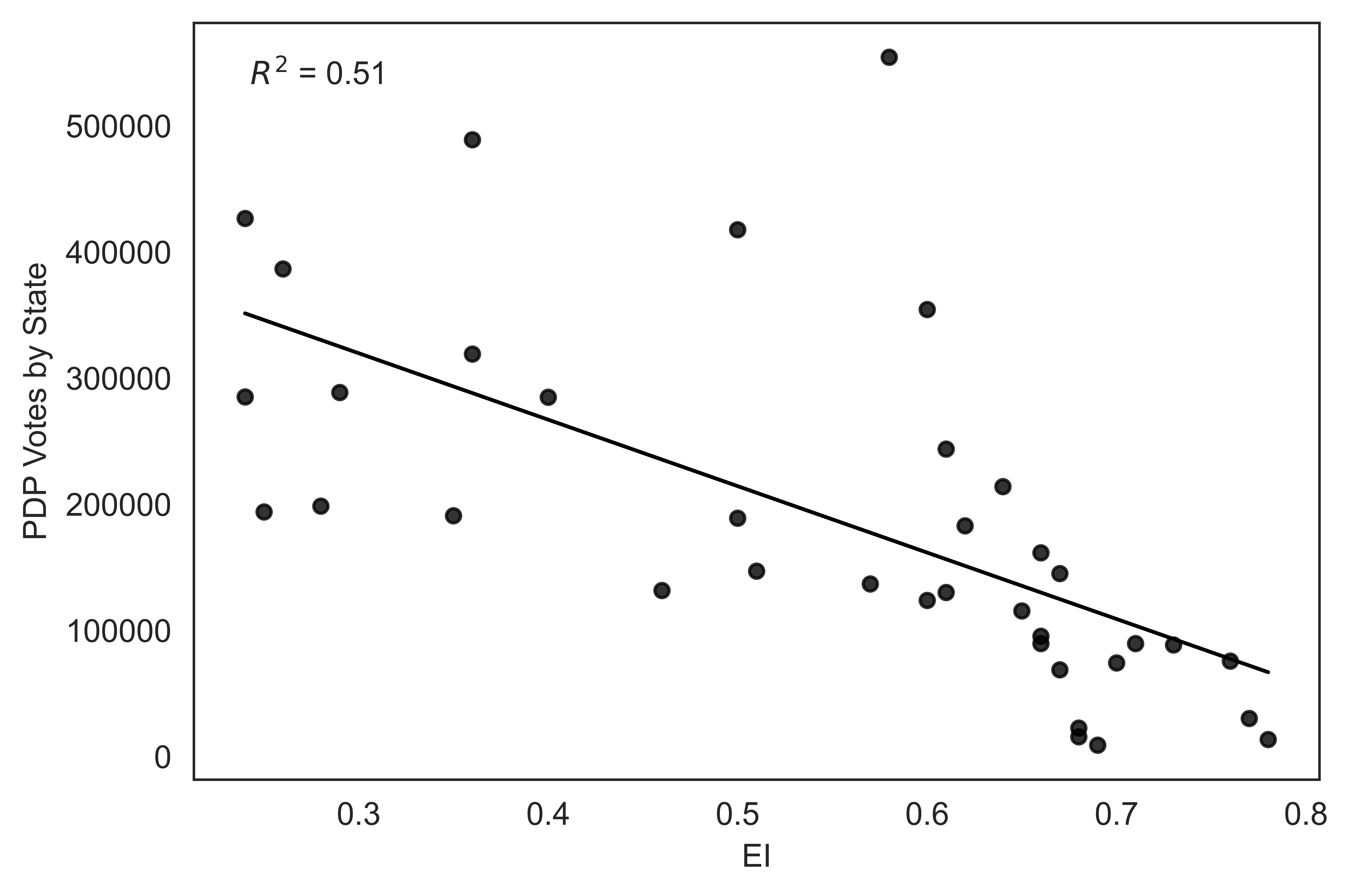}
		\caption{}\label{fig:pdp-ei}
	\end{subfigure}\hfill
	\begin{subfigure}{0.5\textwidth}
		\includegraphics[width=\linewidth]{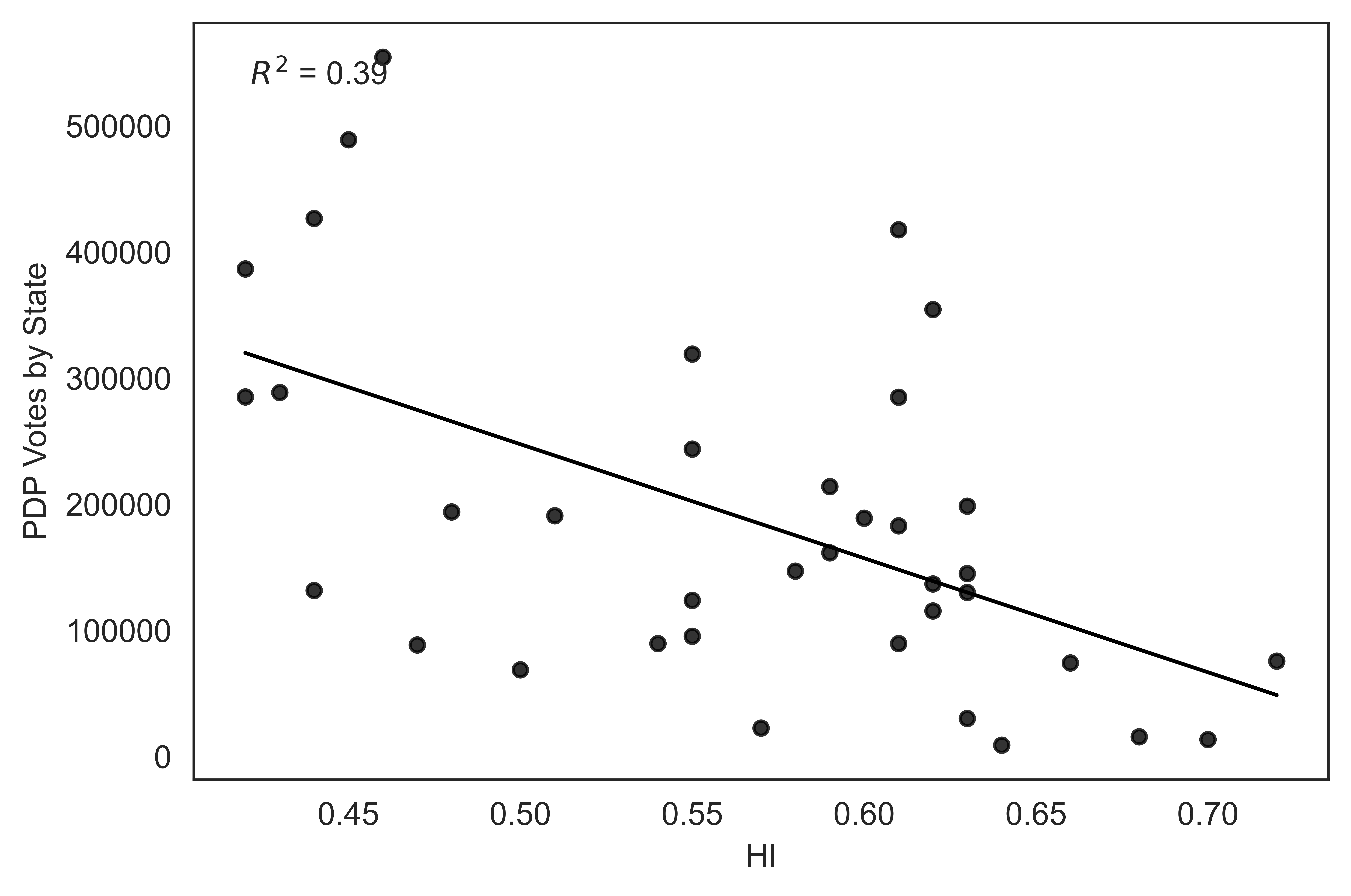}
		\caption{}\label{fig:pdp-hi}
	\end{subfigure}
	
	\vspace{0.5em}
	
	\begin{subfigure}{0.5\textwidth}
		\includegraphics[width=\linewidth]{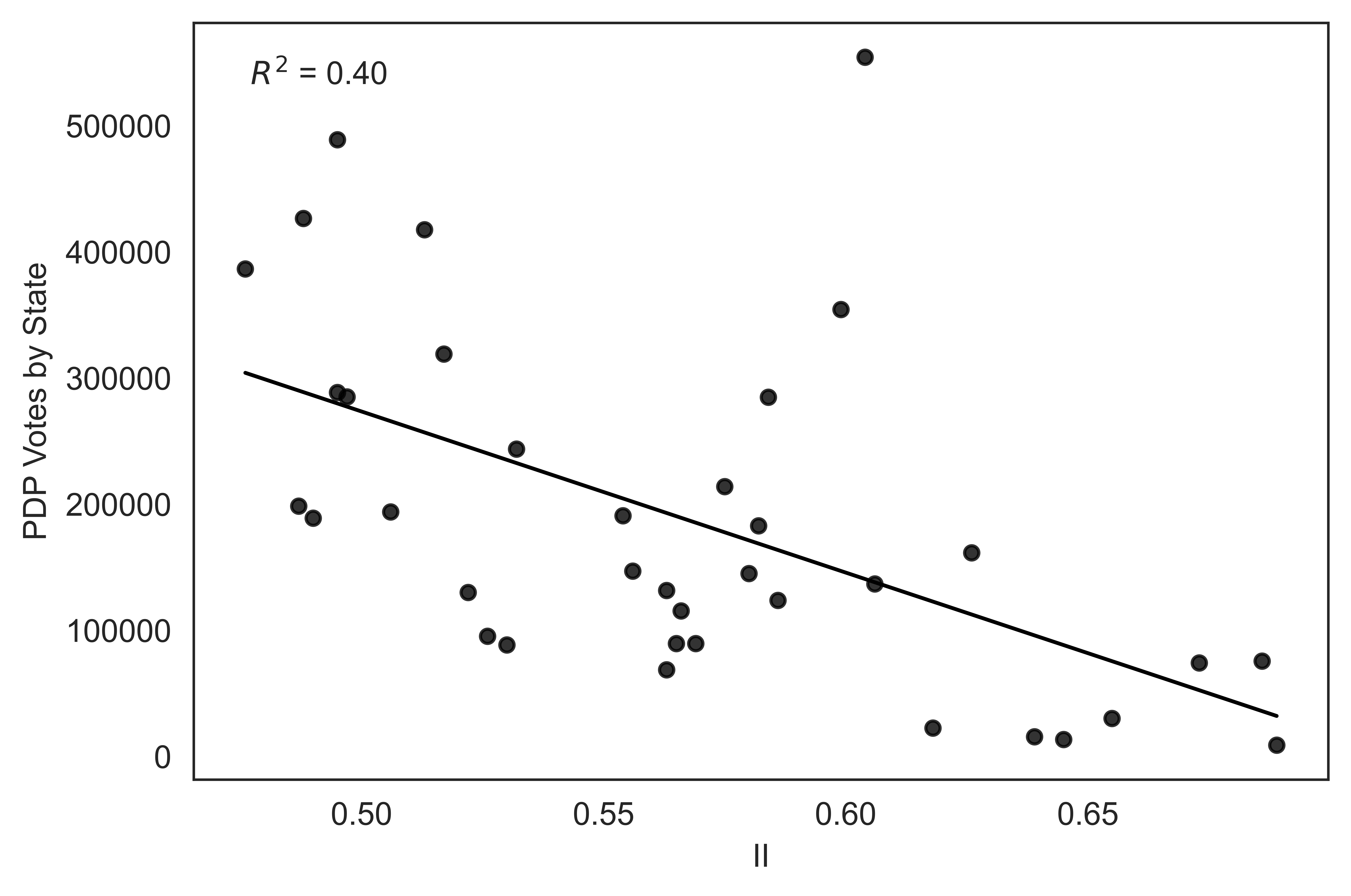}
		\caption{}\label{fig:pdp-ii}
	\end{subfigure}\hfill
	\begin{subfigure}{0.5\textwidth}
		\includegraphics[width=\linewidth]{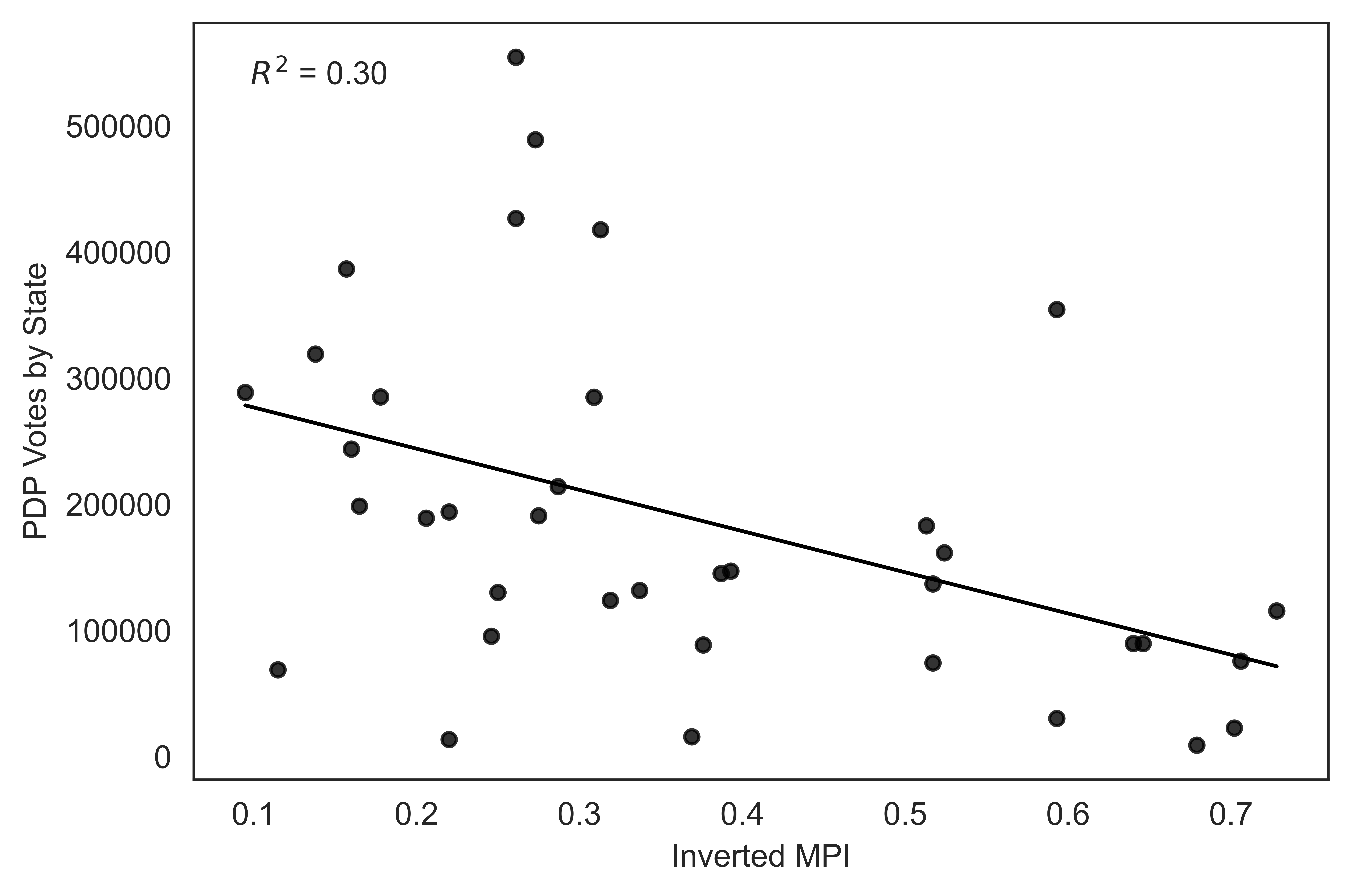}
		\caption{}\label{fig:pdp-invmpi}
	\end{subfigure}
	
	\caption{PDP votes versus socioeconomic indicators}
	\label{fig:pdp-indicators}
\end{figure}

In contrast to the LP, the APC in Figure~\ref{fig:apc-indicators} exhibits a generally negative 
relationship with the same socioeconomic indicators. Figures~\ref{fig:apc-ei} and \ref{fig:apc-hi} show declining trends in APC votes as educational attainment and health outcomes 
improve, with these indicators explaining about 21\% and 20\% of the variation 
respectively. This suggests that the party received relatively less support in 
states with high EI and HI. Figures~\ref{fig:apc-ii} and ~\ref{fig:apc-invmpi}, which shows II and the well-being index respectively, also display weak and negative associations with low explaining power indicating that other factors such as ethno-religious identity, and regional loyalties may have contributed to the unexplained variation in APC voting behaviour. Also, these patterns suggest that APC support in the 2023 election was relatively stronger in less-developed areas, highlighting an inverse relationship between socioeconomic development and voting outcomes compared to the LP. Finally, in Figure~\ref{fig:pdp-indicators} the socioeconomic voting patterns for the PDP is examined, offering further insight into how development indicators shaped electoral support across the three leading political parties.

	\FloatBarrier
The PDP in Figures~\ref{fig:pdp-indicators}, show the relationship between PDP votes across states and the socioeconomic indicators, revealing consistent negative associations, with varying degrees of explanatory power. These patterns suggest that the PDP drew more support from states with low socioeconomic metrics in the 2023 presidential election. In Figure~\ref{fig:pdp-ei}, the EI explains approximately 51\% of the variation in PDP votes implying a moderate strong explanatory power but in negative direction, indicating that states with lower educational attainment were more likely to support the PDP. Figure~\ref{fig:pdp-hi} shows that HI accounts for about 47\% of the variation but in negative direction, the same pattern is equally observed in Figure~\ref{fig:pdp-ii} and Figure~\ref{fig:pdp-invmpi} further suggesting that states with low socioeconomic indices tended to favor the PDP. Overall, these findings suggest that socioeconomic underdevelopment played a significant role in shaping PDP voter support, similar to the pattern observed for the APC. In contrast to the LP, which found greater resonance in more developed states. This divergence highlights a clear polarization in political alignment, with socioeconomic status serving as a key driver of electoral behavior across party lines. These explanatory percentages across the LP, APC, and PDP quantifies the extent to which the socioeconomic indicators explain voting support for each of the parties thereby addressing RQ2. To further understand the impact of socioeconomic factors on the wining probability of each party, the MNL model for predicting party winning probability is presented in the subsequent section.

\subsection{Predicting Party Win Probability based on the Socioeconomic Indicator}
\begin{figure}[htbp]
	\centering
	
	\begin{subfigure}[t]{0.65\linewidth}
		\centering
		\includegraphics[width=\linewidth]{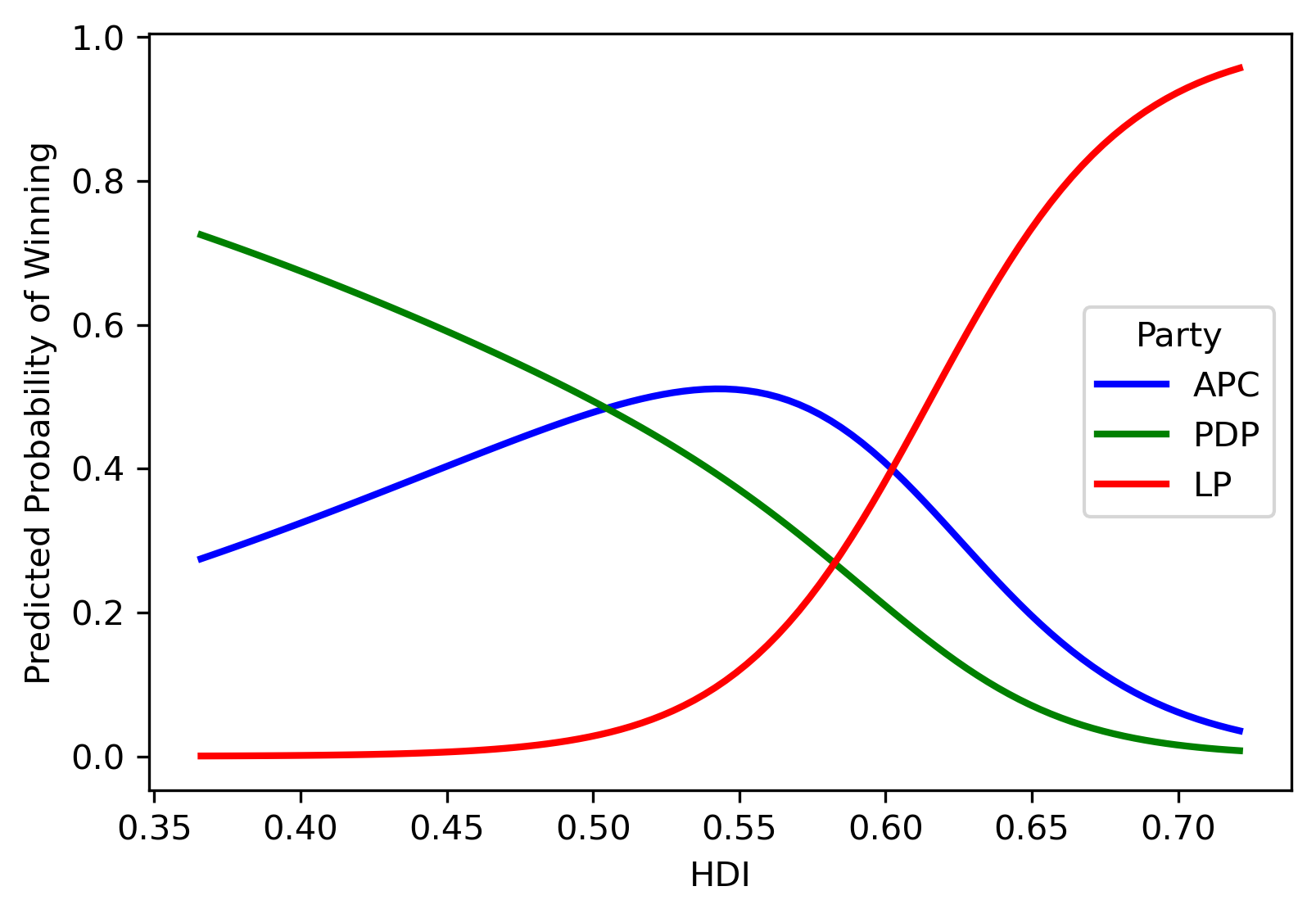} 
		\caption{}
		\label{fig:(aa)}
	\end{subfigure}\hfill
	%
	\begin{subfigure}[t]{0.65\linewidth}
		\centering
		\includegraphics[width=\linewidth]{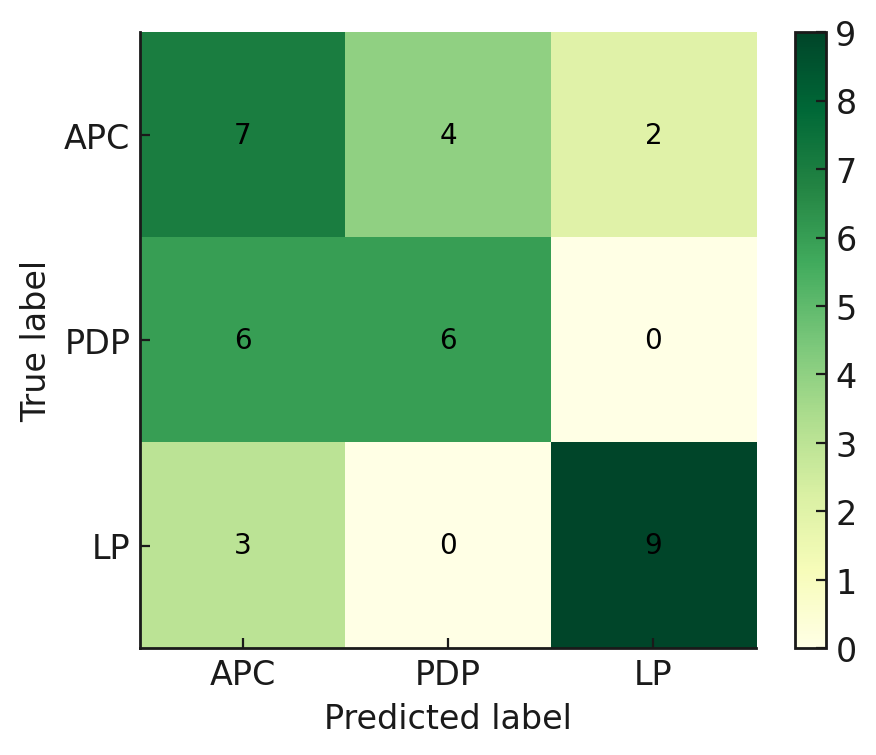} 
		\caption{}
		\label{fig:(bb)}
	\end{subfigure}
	
	\caption{(a) Predicted probabilities of APC, PDP, and LP winning as a function of HDI (b) Confusion matrix of predicted vs. actual state-level winners}
	\label{fig:fig22}
\end{figure}

The MNL empirical predictors model for party-specific support based on HDI of the states in the 2023 election, with APC as the reference category, is specified in Equations~\ref{eq:mnl-lp} to \ref{eq:mnl-apc}, and the corresponding party winning probabilities are given in Equations~\ref{eq:mnl-prob-lp} to \ref{eq:mnl-prob-apc}. Results for the 36 states and the F.C.T. are presented in Table~\ref{tab:state_probs}, which compares HDI-based predictions with actual winners. The trends and confusion matrix of predicted versus true labels are shown in Figure~\ref{fig:fig22}, while model precision is presented in Table~\ref{tab:classification_report}, thereby addressing RQ3.

\begin{align}
	\eta_{\text{LP},s}  &= –16.7094 +27.7470\,\text{HDI}_s, \label{eq:mnl-lp}\\
	\eta_{\text{PDP},s} &= +3.5349 - –7.0054\,\text{HDI}_s, \label{eq:mnl-pdp}\\
	\eta_{\text{APC},s} &= 0. \label{eq:mnl-apc}
\end{align}

\begin{align}
Win	Pr(y_s=\text{LP})  &= 
	\frac{\exp(\eta_{\text{LP},s})}
	{\exp(\eta_{\text{LP},s}) + \exp(\eta_{\text{PDP},s}) + 1}, \label{eq:mnl-prob-lp}\\[2pt]
Win	Pr(y_s=\text{PDP}) &= 
	\frac{\exp(\eta_{\text{PDP},s})}
	{\exp(\eta_{\text{LP},s}) + \exp(\eta_{\text{PDP},s}) + 1}, \label{eq:mnl-prob-pdp}\\[2pt]
Win	Pr(y_s=\text{APC}) &= 
	\frac{1}{\exp(\eta_{\text{LP},s}) + \exp(\eta_{\text{PDP},s}) + 1}. \label{eq:mnl-prob-apc}
\end{align}

\vspace*{-2em}

\begin{table}[H]
	\centering
	\scriptsize
	\setlength{\tabcolsep}{4pt}
	\caption{State-level HDI and predicted vs. actual winning probabilities in the 2023 presidential election. The (*) implies placeholder.}
	\label{tab:state_probs}
	\begin{tabular}{l c c c c l l}
		\toprule
		State & HDI & Pr(APC) & Pr(LP) & Pr(PDP) & Predicted\_Winner & Actual\_Winner \\
		\midrule
		Abia & 0.622 & 32\% & 55\% & 14\% & LP & LP \\
		Adamawa & 0.539 & 51\% & 9\% & 40\% & APC & PDP \\
		Akwa Ibom & 0.602 & 40\% & 40\% & 20\% & APC & PDP \\
		Anambra & 0.674 & 12\% & 85\% & 4\% & LP & LP \\
		Bauchi & 0.372 & 28\% & 0\% & 72\% & PDP & PDP \\
		Bayelsa & 0.573 & 48\% & 22\% & 30\% & APC & PDP \\
		Benue & 0.582 & 46\% & 27\% & 27\% & APC & APC \\
		Borno & 0.464 & 43\% & 1\% & 57\% & PDP & APC \\
		Cross River & 0.575 & 48\% & 23\% & 29\% & APC & LP \\
		Delta & 0.607 & 38\% & 43\% & 19\% & LP & LP \\
		Ebonyi & 0.706 & 5\% & 93\% & 1\% & LP & LP \\
		Edo & 0.633 & 27\% & 63\% & 11\% & LP & LP \\
		Ekiti & 0.612 & 36\% & 47\% & 17\% & LP & APC \\
		Enugu & 0.667 & 14\% & 82\% & 4\% & LP & LP \\
		F.C.T & 0.678 & 11\% & 86\% & 3\% & LP & LP \\
		Gombe & 0.466 & 43\% & 1\% & 56\% & PDP & PDP \\
		Imo & 0.683 & 9\% & 88\% & 3\% & LP & LP \\
		Jigawa & 0.371 & 28\% & 0\% & 72\% & PDP & APC \\
		Kaduna & 0.545 & 51\% & 10\% & 38\% & APC & PDP \\
		Kano & 0.482 & 45\% & 2\% & 53\% & PDP* & APC* \\
		Kastina & 0.431 & 37\% & 0\% & 62\% & PDP & PDP \\
		Kebbi & 0.366 & 27\% & 0\% & 73\% & PDP & PDP \\
		Kogi & 0.625 & 30\% & 57\% & 13\% & APC & APC \\
		Kwara & 0.597 & 42\% & 36\% & 22\% & APC & APC \\
		Lagos & 0.721 & 4\% & 96\% & 1\% & LP & LP \\
		Nasarawa & 0.549 & 51\% & 12\% & 37\% & APC & LP \\
		Niger & 0.523 & 50\% & 6\% & 44\% & APC & APC \\
		Ogun & 0.579 & 47\% & 25\% & 28\% & APC & APC \\
		Ondo & 0.611 & 46\% & 36\% & 17\% & APC & APC \\
		Osun & 0.607 & 43\% & 38\% & 19\% & APC & PDP \\
		Oyo & 0.603 & 41\% & 39\% & 20\% & APC & APC \\
		Plateau & 0.563 & 50\% & 17\% & 33\% & APC & LP \\
		Rivers & 0.601 & 40\% & 39\% & 21\% & LP & APC \\
		Sokoto & 0.397 & 32\% & 0\% & 68\% & PDP & PDP \\
		Taraba & 0.527 & 51\% & 6\% & 43\% & APC & PDP \\
		Yobe & 0.439 & 39\% & 0\% & 61\% & PDP & PDP \\
		Zamfara & 0.392 & 31\% & 0\% & 69\% & PDP & APC \\
		\bottomrule
	\end{tabular}
\end{table}

\FloatBarrier

Figure~\ref{fig:fig22} highlights the relationship between human development and electoral outcomes, as well as the predictive performance of the multinomial logit model. Figure~\ref{fig:(aa)} shows that the probability of LP winning rises steadily with higher HDI values, becoming dominant above an HDI of 0.65. In contrast, APC dominates at lower HDI levels (below ~0.50), while PDP is most competitive in the mid-HDI range (around 0.50–0.60), indicating that party support is patterned by levels of socioeconomic development. Figure~\ref{fig:(bb)} displays the confusion matrix comparing predicted versus actual state-level winners. The model demonstrates the highest accuracy for LP, correctly classifying 9 out of 12 states, while APC and PDP show more misclassification — each with 6 incorrect predictions. This suggests the model performs best at identifying LP victories and struggles more in distinguishing between APC and PDP. These findings are consistent with the precision, recall, and F1-scores reported in Table~\ref{tab:classification_report}, which reflect stronger predictive performance for LP. Overall, the model not only provides a reasonable level of predictive accuracy but also offers insights into the structural association between human development and electoral competitiveness across parties.
\vspace{-2.0em}
\FloatBarrier

\begin{table}[htbp]
	\centering
	\scriptsize
	\setlength{\tabcolsep}{6pt}
	\caption{Classification report showing precision, recall, and F1-scores for predicted party outcomes based on HDI (2023 presidential election).}
	\label{tab:classification_report}
	\begin{tabular}{lcccc}
		\toprule
		Party & Precision & Recall & F1-score & Support \\
		\midrule
		APC & 0.438 & 0.538 & 0.483 & 13 \\
		PDP & 0.600 & 0.500 & 0.545 & 12 \\
		LP  & 0.818 & 0.750 & 0.783 & 12 \\
		\midrule
		Accuracy   &  &  & 0.595 & 37 (60\%) \\
		Macro avg  & 0.619 & 0.596 & 0.604 & 37 \\
		Weighted avg & 0.614 & 0.595 & 0.600 & 37 \\
		\bottomrule
	\end{tabular}
\end{table}

\subsection{Comparing HDI–Vote Correlation and Macroeconomic Post-Election Across Election Cycles}
\paragraph{} 
Table~\ref{tab:csds_compact} presents a cross-year comparison of Nigerian presidential elections from 2011 to 2023, showing how the top 10 most socioeconomically developed states (as measured by HDI) voted, the correlation between HDI and each party’s vote share, the eventual electoral winner, and the average macroeconomic performance post-election. In 2023, the LP showed a strong positive correlation (\(0.66\)) with HDI and won 8 of the 10 states with highest socioeconomic indices, although the APC, which had a negative correlation (\(-0.34\)), ultimately won the presidency. Post-election, macroeconomic performance was modest, with rebased average GDP at 3.22\%, unemployment at 3.37\%, and inflation high at 26.10\%.

\paragraph{}
In earlier cycles (2019 and 2015), the APC also won despite negative correlations (\(-0.43\), \(-0.38\)) with HDI indicating support from from states with low socioeconomic indices. The post-election years were marked by weaker GDP growth and relatively high unemployment and inflation, suggesting that administrations not voted massively by states with high socioeconomic indices tend to underperform economically, although further study is needed to make a definite conclusion. In contrast, the 2011 election saw the PDP win with a strong positive correlation (\(0.44\)) and capture 9 of the 10states with highest socioeconomic indices. This period also recorded the most favorable macroeconomic outcomes, with 5.63\% GDP growth, 3.92\% unemployment, and 9.90\% inflation. Comparatively, the data suggest that stronger alignment between electoral support and more developed states may be associated with better macroeconomic outcomes post-election. Conversely, when the electoral mandate is driven by states with low HDI and negative correlation which addresses RQ4. This pattern supports the idea that development-aligned mandates may foster stronger macroeconomic performance.
\begin{table}[htbp]
	\centering
	\scriptsize
	\setlength{\tabcolsep}{3pt}      
	\renewcommand{\arraystretch}{0.9} 
	\caption{HDI–Vote Correlation and Macroeconomic Outcomes Post-Elections for 2011-2023 Election Cycle}
	\label{tab:csds_compact}
	\begin{tabular}{@{}c l c c l c c c@{}}
		\toprule
		\textbf{Year} & \textbf{Party} & \textbf{Top-10 States} & \textbf{Corr.}
		& \textbf{Winner} & \textbf{GDP \%} & \textbf{Unemp \%} & \textbf{Infl \%} \\
		\midrule
		\multirow{3}{*}{2023}
		& APC & 2 & $-0.34$ & APC & 3.22 & 3.37 & 26.10 \\
		& LP  & 8 & $+0.66$ &     &      &      &       \\
		& PDP & 0 & $-0.68$ &     &      &      &       \\
		\midrule
		\multirow{3}{*}{2019}
		& APC    & 4 & $-0.43$ & APC & 1.83 & 7.86 & 15.11 \\
		& PDP    & 6 & $+0.49$ &     &      &      &       \\
		& Others & 0 & $+0.39$ &     &      &      &       \\
		\midrule
		\multirow{3}{*}{2015}
		& APC    & 6 & $-0.38$ & APC & 0.95 & 7.00 & 13.33 \\
		& PDP    & 4 & $+0.37$ &     &      &      &       \\
		& Others & 0 & $+0.22$ &     &      &      &       \\
		\midrule
		\multirow{3}{*}{2011}
		& ACN & 1 & $+0.30$ &     &      &      &      \\
		& CPC & 2 & $-0.50$ &     &      &      &      \\
		& PDP & 7 & $+0.44$ & PDP & 5.63 & 3.92 & 9.90 \\
		\bottomrule
	\end{tabular}
	\vspace{-4pt}
\end{table}
\FloatBarrier

\section{Discussion}

Understanding the dynamics of voter behavior in contemporary democracies—particularly in a developing country like Nigeria—requires looking beyond conventional narratives to examine the underlying socioeconomic drivers that shape electoral outcomes. At the same time, it is essential to account for other factors such as vote buying, electoral intimidation, voter suppression, manipulation of results, weak institutions, media influence, and the persistence of ethno-religious and regional loyalties, all of which continue to interact with socioeconomic conditions in shaping how citizens make electoral choices.The 2023 Nigerian presidential election demonstrates how socioeconomic conditions are increasingly shaping electoral outcomes alongside traditional identity-based cleavages. While ethnicity and religion remain powerful drivers of political loyalty, this study reveals that states with higher levels of education index, income index, health index, and low multidimensional poverty  were more likely to support reformist candidates perceived to be a technocrat, less corrupt, engaging, and resonate with the youth and educated demography. In contrast, regions with lower socioeconomic indices gravitated toward the established politicians in dominant parties. This divergence underscores a political polarization that reflects the uneven geography of development across the federation, supporting the findings by  \citet{IdowuIyabode2024} on entrenched identity politics.

The implications of these results are significant for Nigeria’s democracy. On one hand, they suggest that issue-based voting is gaining traction as noted by \citet{Olabanjo2023} and \citet{Oyewola2023}, who highlight the growing influence of digital engagement especially on social media in shaping voter behavior, particularly in areas where citizens enjoy greater socioeconomic empowerment. Educated and economically secure populations may be more inclined to evaluate candidates on governance capacity, reformist credentials, and developmental promises. ~\citet{Nwangbo2024} noted the high level of engagement by the LP candidate, particularly among youths through social media, during the 2023 electioneering. This mobilization may explain why voters in many states with higher socioeconomic indices tilted toward the LP candidate, despite his position as an underdog in the contest. Additionally, the persistence of support for established parties in states with low socioeconomic indices highlights the continuing role of clientelism, regional loyalties, identity politics, and weaponization of poverty to gain political patronages by entrenched politicians as emphasized by \citet{Salahu2023}. These dynamics reveal that Nigerian democracy is caught in a transitional moment between entrenched patronage systems and emerging issue-based contestation.

Candidates who can articulate credible visions for education, job creation, and poverty reduction may find new leverage among upwardly mobile electorates. Although \citet{Babalola2024} argues that political parties in Nigeria largely function as instruments for elite bargaining with little or no ideological foundation, this study support that view. Political party ideology, as important as it is, has remained weak not only in Nigeria but across much of the African countries. Party platforms often remain vague and lack clear programmatic commitments. Nevertheless, individual candidates can still embody ideological narratives—particularly around governance, reform, and anti-corruption—that resonate with segments of the enlightened electorate, thereby filling the gap left by weak party structures. In this case, the LP candidate’s appeal during 2023 presidential election suggests that ideology, while weakly institutionalized, can nonetheless emerge through candidate perception and public discourse.

Finally, these findings highlight important questions about the complex relationship between development and the quality of democracy. In more developed contexts, electorates with greater access to information and resources are often better positioned to support reform-oriented candidates, creating the potential for stronger alignment between governance capacity and voter expectations. However, the continued influence of less-developed regions in shaping national outcomes underscores the persistent structural barriers to consolidating accountable democracy in societies marked by deep inequality.

\section{Conclusion}

The 2023 Nigerian presidential election, often framed around ethno-religious affiliation and patronage, also revealed the influence of socioeconomic factors on voter behavior. This study explored whether development indicators—education, income, health, and poverty—shaped electoral outcomes. Voter behavior across states displayed clear dynamism based on the socioeconomic factors when assessed through spatio-temporal analysis. Using a multinomial logit model, the probability of a party winning a state based on the HDI was predicted with high accuracy, correctly forecasting more than two-thirds of the outcomes and highlighting its value for electoral studies. Spearman correlation further showed that stronger positive associations between state-level HDI and votes for the eventual winner were consistently associated to macroeconomic performance during the tenure of the winners, a trend observed across the presidential electoral cycles from 2011 to 2023. Additionally, the findings show that support for the LP was strongest in the states with high socioeconomic indices compared the APC and PDP. Therefore, apart from ethno-religious sentiments, socioeconomic development plays a significant role in shaping electoral preferences in Nigeria especially among educated and younger demographics.

Despite the strengths of the empirical approach, this study has notable limitations. Reliance on state-level aggregates may obscure important intra-state variations, while the cross-sectional design limits the ability to capture dynamic shifts in voter sentiment, such as late campaign swings. Although multiple electoral cycles were analyzed, the design may not fully reflect within-cycle changes. Furthermore, while strong associations between HDI and voting outcomes were identified, causality cannot be established with certainty. Future research should statistically test whether with stronger alignment between electoral support and states with higher socioeconomic indices is associated with improved macroeconomic outcomes with certainty across a wider range of election cycles. From a policy perspective, government should prioritize strengthening socioeconomic conditions particularly in education, healthcare, and poverty reduction not only to improve citizens’ well-being but also to cultivate a more informed electorate capable of supporting competent, reform-oriented leaders. In the long run, such efforts could foster a more accountable and development-focused democratic process.

\subsubsection*{Conflicts of Interest} 
The authors declare no conflicts of interest

\bibliographystyle{elsarticle-harv} 
\bibliography{refs}                 

\end{document}